\documentclass[useAMS,usenatbib,longnamesfirst,usedcolumn]{mn2e} 
\usepackage{epsfig}	
\usepackage{times}	
\title[Temperature \& Cooling time profiles]{A statistically-selected Chandra 
 sample of 20 galaxy clusters -- I. Temperature and cooling time profiles}
\author[A. J. R. Sanderson, T. J. Ponman and E. O'Sullivan]
       {Alastair J. R. Sanderson$^{1,2}$ \thanks{E-mail: ajrs@star.sr.bham.ac.uk},
        Trevor J. Ponman$^{1}$ and Ewan O'Sullivan$^{3}$\\
 $^{1}$School of Physics and Astronomy, University of
        Birmingham, Edgbaston, Birmingham B15 2TT, UK \\
 $^{2}$Department of Astronomy, University of Illinois, 
        1002 West Green Street, Urbana, IL 61801, USA \\
 $^{3}$Harvard-Smithsonian Center for Astrophysics, 60 Garden Street,
        Cambridge, MA 02138\\
       \\}
 \date{Accepted 2006 August 18.
      Received 2006 August 18;
      in original form 2006 May 24 ($svn$ $Revision: 71 $)}

\pagerange{\pageref{firstpage}--\pageref{lastpage}}
\pubyear{2006}

\shortcites{allen01b,birzan04,bauer05,blanton04,buote03,cortese04,
 donahue06,ensslin98,finoguenov01,ghizzardi05,girardi97,gomez02,ikebe02,
 johnstone02,ledlow03,mar98b,mccarthy04,mcnamara05,motl04,ohara06,osu05a,
 peres98,piffaretti05,poole06,rottgering94,rottgering97,san03,slee01,
 sun03,sun02b,venturi02,vikhlinin05,voit03b,werner06,willis05,zhang06}

\newcommand{\rmsub}[2]{\ensuremath{#1_{\mathrm{#2}}}} 
\newcommand{\srel}[2]{\mbox{\ensuremath{#1 - #2}}} 
\newcommand{\ASCA}{\textit{ASCA}}

\newcommand{\Chandra}{\textit{Chandra}}

\newcommand{\CIAO}{\textsc{ciao}}

\newcommand{\eg}{{\textrm e.g.}}

\newcommand{\km}{\ensuremath{\mbox{~km}}}
\newcommand{\kmpspMpc}{\ensuremath{\km \ps \pMpc\,}}

\newcommand{\LX}{\rmsub{L}{X}}
\newcommand{\Mpc}{\ensuremath{\mbox{~Mpc}}}

\newcommand{\MT}{\srel{M}{\TX}}
\newcommand{\omegal}{\rmsub{\Omega}{\Lambda}}
\newcommand{\omegam}{\rmsub{\Omega}{m}}

\newcommand{\pMpc}{\ensuremath{\Mpc^{-1}}}
\newcommand{\ps}{\ensuremath{\s^{-1}}}
\newcommand{\rtwoh}{\rmsub{r}{200}}
\newcommand{\rfiveh}{\rmsub{r}{500}}

\newcommand{\rhogas}{\rmsub{\rho}{gas}}
\newcommand{\ROSAT}{\textit{ROSAT}}
\newcommand{\Rproject}{\textsc{r project}}

\newcommand{\s}{\ensuremath{\mbox{~s}}}
\newcommand{\Tbar}{\ensuremath{\overline{T}}}
\newcommand{\tcool}{\rmsub{t}{cool}}
\newcommand{\TX}{\rmsub{T}{X}}

\newcommand{\XMM}{\emph{XMM-Newton}}
\newcommand{\XSPEC}{\textsc{xspec}}
\begin{document}

\maketitle

\label{firstpage}

\begin{abstract}
  
 \noindent

 We present an analysis of 20 galaxy clusters observed with the \Chandra\
 X-ray satellite, focussing on the temperature structure of the
 intracluster medium and the cooling time of the gas. Our sample is drawn
 from a flux-limited catalogue but excludes the Fornax, Coma and Centaurus
 clusters, owing to their large angular size compared to the \Chandra\
 field-of-view. We describe a quantitative measure of the impact of central
 cooling, and find that the sample comprises 9 clusters possessing cool
 cores and 11 without. The properties of these two types differ markedly,
 but there is a high degree of uniformity amongst the cool core clusters,
 which obey a nearly universal radial scaling in temperature of the form
 $T\propto r^{\sim0.4}$, within the core. This uniformity persists in the
 gas cooling time, which varies more strongly with radius in cool core
 clusters ($\tcool \propto r^{\sim1.3}$), reaching $\tcool <1\,$Gyr in all
 cases, although surprisingly low central cooling times ($<5\,$Gyr) are
 found in many of the non-cool core systems. The scatter between the
 cooling time profiles of all the clusters is found to be remarkably small,
 implying a universal form for the cooling time of gas at a given physical
 radius in virialized systems, in agreement with recent previous work. Our
 results favour cluster merging as the primary factor in preventing the
 formation of cool cores.

\end{abstract}



\section{Introduction}
\label{sec:intro}

Clusters of galaxies are uniquely suitable environments for studying galaxy
formation and evolution, as well as excellent cosmological tools for
measuring the fundamental parameters of the Universe. A study of the
scaling properties of clusters, which span roughly two decades in mass, is one
of the best means of tackling both these objectives, and many analyses have
been conducted on this basis, establishing broad trends and quantifying
departures from simple self-similarity
\citep[e.g.][]{edge91,mar98b,hel00,fukazawa04}.

Increasingly, however, attention is shifting towards a more complete
understanding of cluster physics, focussing in particular on the scatter in
these scaling relations. With current generation telescopes, we are
reaching the point where the precision of scaling relations is limited by
systematic effects rather than measurement errors. Further progress
requires a better knowledge of cluster physics to understand the sources of
intrinsic scatter in these relations. Not only does this hold great
potential for understanding galaxy evolution, it also promises to hone
cluster observables as proxies for mass, which will be vital in extracting
precise cosmological constraints from forthcoming large cluster surveys.

Since the cooling time of gas in the centres of most clusters is short
compared to their age, radiative cooling would be expected to play an
important role in cluster physics. However, many clusters do not show
strong signs of cooling, possibly as a result of disruption caused by
merging and associated shock heating of the gas, or else non-gravitational
energy input (e.g. from AGN), or even heat redistribution via conduction or
mixing. Moreover, even where cooling dominates, its progress is apparently
impeded \citep[see the recent review by][and references
therein]{peterson06}, by processes which may well relate to these same
phenomena. Whatever the explanation, it is clear that clusters can be
separated into two distinct categories, according to the presence or
absence of a `cool core' \citep{peres98,bauer05}, where radiative heat
losses have significantly lowered the central temperature of the gas. It is
the combination of cool core and non cool core systems in the cluster
population which accounts for much of the intrinsic scatter in scaling
relations.

While cool core clusters have tended to be relatively well-studied,
non-cool core clusters have received less attention in scaling relation
analyses. Consequently there is great potential for exploring cluster
physics by comparing the properties of the two types of clusters. This aim
of this work is to investigate precisely this comparison, by focussing on a
systematically selected sample of clusters, using data from the \Chandra\
satellite, to allow a high-resolution study of the core properties of the
intracluster medium (ICM). 

Previous detailed studies of the temperature structure of the hot gas in
clusters have generally concentrated on the most relaxed
(i.e. morphologically regular) clusters
\citep[e.g.][]{san03,piffaretti05,vikhlinin05}, which frequently possess
cool cores. In this work, we have compiled a statistically selected sample
of 20 clusters, to provide a more rigorous selection process. This
naturally includes a larger number of non-cool core and perhaps
less-relaxed clusters, which are more representative of the cluster
population as a whole. We have subjected these 20 systems to a
non-parametric analysis, so as to avoid any model-dependent biases in the
comparison of cool core (CC) and non-CC systems. Furthermore, in
restricting the analysis to X-ray data taken with the \Chandra\ satellite,
we are able to exploit the unrivalled spatial resolution of this telescope,
to permit a detailed examination of the inner regions of the ICM, where the
impact of radiative cooling is most pronounced.

Throughout this paper we adopt the following cosmological parameters;
$H_{0}=70$\kmpspMpc, $\rmsub{\Omega}{m}=0.3$ and
$\Omega_{\Lambda}=0.7$. Throughout our spectral analysis we have used \XSPEC\
11.3.1, incorporating the solar abundance table of \citet{grevesse98},
which is different from the default abundance table. Typically this results
in larger Fe abundances, by a factor of $\sim$1.4. All errors are
1$\sigma$, unless otherwise stated.

\section{Sample Selection and Properties}
The sample comprises the 20 highest flux clusters drawn from the 63
cluster, flux-limited sample of \citet{ikebe02}, excluding those objects
with extremely large angular sizes (the Coma, Fornax and Centaurus
clusters), which are difficult to observe with \Chandra\ owing to its
limited field of view. Some basic properties of this sample are listed in
Table~\ref{tab:sample}. The \citeauthor{ikebe02} flux-limited sample was
constructed from the HIFLUGS sample of \citet{reiprich02}, additionally
selecting clusters lying above an absolute galactic latitude of 20 degrees
and located outside of the Magellanic Clouds and the Virgo Cluster regions.

\begin{figure}
\includegraphics[width=8cm]{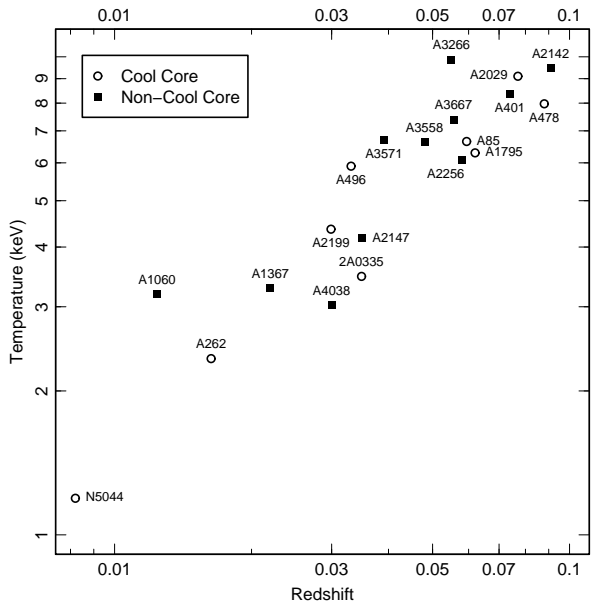}
\caption{ \label{fig:z-kT}
  Cluster mean temperature within 0.1--0.2 \rfiveh\ as a function of
  redshift, labelled according to the presence of a cool core. Temperature
  errors have been omitted for clarity. Redshifts are taken from the NASA
  Extragalactic Database (NED).}
\end{figure}

Fig.~\ref{fig:z-kT} shows the cluster redshift as a function of mean
X-ray temperature, excluding any cool core (see
Section~\ref{sec:mean_kT}). The temperature errors are generally too small
to be seen, and have been omitted for clarity. The point style
differentiates between those clusters possessing a significant cool core
and those without. The cool core classification scheme is described in 
Section~\ref{sec:CC}. 

%
%
\begin{table*}
\begin{tabular}{l*{9}{c}}
\hline
Name & Obsid$^{a}$ & Detector$^{b}$ & RA & Dec. & Redshift & HI Column$^{c}$ & Mean kT$^{d}$ & \rfiveh\ & Notes$^{e}$ \\
     &  & & (J2000) & (J2000) &  & ($\times10^{20}$ cm$^{-2}$) & (keV) & (kpc) & \\
\hline\hline\\[-2ex] 
  NGC 5044 &  798 &  S & 198.850 & -16.385 & 0.008 & 4.9 & 1.19 $ ^{+0.03}_{-0.02} $ & 436 & -- \\
   Abell 262 & 2215 &  S (VF) &  28.194 &  36.152 & 0.016 & 5.4 & 2.34 $ ^{+0.06}_{-0.06} $ & 668 & -- \\
  Abell 4038 & 4188 &  I (VF) & 356.928 & -28.143 & 0.030 & 1.6 & 3.02 $ ^{+0.17}_{-0.17} $ & 784 & F, T \\
  Abell 1060 & 2220 &  I (VF) & 159.181 & -27.526 & 0.012 & 4.8 & 3.19 $ ^{+0.07}_{-0.07} $ & 812 & F, T \\
  Abell 1367 &  514 &  S & 176.191 &  19.699 & 0.022 & 2.3 & 3.28 $ ^{+0.11}_{-0.10} $  & 826 & T \\
 2A0335+096 &  919 &  S &  54.669 &   9.970 & 0.035 & 17.9 & 3.47 $ ^{+0.14}_{-0.12} $ & 856 & -- \\
  Abell 2147 & 3211 &  I (VF) & 240.567 &  15.963 & 0.035 & 3.4 & 4.17 $ ^{+0.14}_{-0.14} $ & 961 & F, T \\
  Abell 2199 &  497 &  S & 247.160 &  39.551 & 0.030 & 0.9 & 4.36 $ ^{+0.13}_{-0.11} $ & 989 & -- \\
   Abell 496 &  931 &  S &  68.408 & -13.262 & 0.033 & 4.6 & 5.91 $ ^{+0.58}_{-0.56} $ & 1200 & -- \\
  Abell 2256 & 1386 &  I & 256.041 &  78.648 & 0.058 & 4.1 & 6.08 $ ^{+0.30}_{-0.31} $ & 1220 & T \\
  Abell 1795 &  493 &  S (VF) & 207.219 &  26.590 & 0.062 & 1.2 & 6.30 $ ^{+0.17}_{-0.17} $ & 1250 & -- \\
  Abell 3558 & 1646 &  S (VF) & 201.987 & -31.496 & 0.048 & 3.9 & 6.64 $ ^{+0.34}_{-0.51} $ & 1290 & T \\
    Abell 85 &  904 &  I &  10.460 &  -9.303 & 0.059 & 3.4 & 6.65 $ ^{+0.14}_{-0.14} $ & 1290 & -- \\
  Abell 3571 & 4203 &  S (VF) & 206.869 & -32.864 & 0.039 & 3.7 & 6.71 $ ^{+0.15}_{-0.42} $ & 1300 & F, T \\
  Abell 3667 &  889 &  I & 303.129 & -56.841 & 0.056 & 4.7 & 7.39 $ ^{+0.27}_{-0.27} $ & 1380 & T \\
   Abell 478 & 1669 &  S &  63.356 &  10.466 & 0.088 & 15.1 & 7.97 $ ^{+0.18}_{-0.18} $ & 1450 & -- \\
   Abell 401 &  518 &  I &  44.739 &  13.578 & 0.074 & 10.5 & 8.35 $ ^{+0.43}_{-0.61} $ & 1490 & T \\
  Abell 2029 & 4977 &  S & 227.734 &   5.745 & 0.077 & 3.0 & 9.11 $ ^{+0.20}_{-0.20} $ & 1570 & -- \\
  Abell 2142 & 1196 &  S & 239.585 &  27.232 & 0.091 & 4.2 & 9.45 $ ^{+0.65}_{-0.65} $ & 1610 & -- \\
  Abell 3266 &  899 &  I (VF) &  67.815 & -61.456 & 0.055 & 1.6 & 9.86 $ ^{+0.53}_{-0.54} $ & 1650 & F, T \\
\hline
\end{tabular}
\caption{
 Some basic properties of the sample, listed in order of increasing temperature. Coordinates are 
 the centroids used in the X-ray analysis. $^{a}$Chandra observation identifier; $^{b}$Denotes 
 either ACIS-I or ACIS-S (observations telemetered in very faint mode are marked with VF); $^{c}$Galactic HI column, interpolated to the cluster centroid using 
 the data of \citet{dic90}; $^{d}$The mean temperature measured between 0.1--0.2 \rfiveh\ (see 
 Section~\ref{sec:mean_kT}); $^{e}$An ``F'' indicates that the galactic absorption was frozen
 at the HI value for the analysis, while ``T'' marks those clusters for which the projected 
 $T(r)$ was used for the 3d $T(r)$. All errors are 68\% confidence.
}
\label{tab:sample}
\end{table*}

\section{Data Reduction}
The data analysis and reduction were performed with version 3.2.2 of the
standard software --- \Chandra\ Interactive Analysis of Observations
(\CIAO\footnote{http://cxc.harvard.edu/ciao/}), incorporating
\textsc{caldb} version 3.1. For all the observations a new level 2 events
file was generated from the level 1 events file downloaded from the
\Chandra\ archive. This ensures that the latest calibration information is
applied uniformly to the data, irrespective of when the observations were
made.

The following procedure was followed for each observation dataset.
According to which CCD chips were used in the observation, a different
light curve was extracted for CCD chips 5 and 7 separately and the
remaining, front-illuminated chips combined. The recommended criteria for
energy extraction and time binning were
used\footnote{http://cxc.harvard.edu/contrib/maxim/bg}. Flares were
identified and excluded with the sigma clipping algorithm implemented in
the \CIAO\ task `lc\_clean', using the median light curve value to provide a
robust initial estimate of the quiescent mean level.

Cosmic ray events were identified and excluded using the \CIAO\ task
`acis\_run\_hotpix', and those found in this manner were extracted
separately and examined to check that photons from the cluster itself were
not misidentified. A new level 1 events file was produced by reprocessing
the resulting events file, to apply the latest gain file. Corrections were
applied for the effects of charge transfer inefficiency (CTI) and
time-dependent gain variation, where necessary. Bad columns and hot pixels
were then excluded, and only events with Advanced Satellite for Cosmology
and Astrophysics (\ASCA) grades 0, 2, 3, 4, and 6 were
retained. Subsequently, a new level 2 events file was generated by
reprocessing this modified level 1 events data set. For those observations
telemetered in very faint (VF) mode (indicated in column~3 of
Table~\ref{tab:sample}), the extra background event flagging and removal
that this enables was performed in both the main and corresponding blank
sky datasets.  

Owing to the close proximity of the clusters in this sample, emission from
the target fills the entire \Chandra\ field of view in most cases. It was
therefore necessary to employ separate background events files for each
observation, using the Markevitch blank-sky data
sets\footnote{http://cxc.harvard.edu/contrib/maxim/acisbg.}. To allow for
small variations in the particle background level between the blank-sky
fields and the target observation, we rescaled the effective exposure of
the background data sets according to the ratio of count rates in the
particle-dominated 7--12 keV energy band. This ratio was calculated for
those CCD chips not considered part of the main detector (i.e. excluding
chips 0-3 for ACIS-I and chip 7 for ACIS-S observations). Generally there
is good agreement between the ratios found for different chips in a given
observation. To avoid the bias caused by contaminating point sources in the
determination of the background rescale factors, we identified and excluded
such features using the iterative method described in \citet{san05}.

To gauge the sensitivity of our results to variations in the normalization
of the blank sky datasets, we investigated its impact on our analysis of
Abell~4038 (mean temperature $\sim$3 keV) --- the cluster most susceptible
to this effect (i.e. where the background is most dominant). We refitted
spectra from our annular profile (see Section~\ref{ssec:spec_prof}),
separately adjusting the background normalization by 10 per cent higher and
lower. In the outermost annulus, this biased the recovered temperature by
roughly 1$\sigma$, in the direction of increasing temperature with higher
background. However, the bias rapidly diminishes for the inner annuli,
dropping below 0.3$\sigma$ for the 3rd innermost spectrum. It is clear,
therefore, that our results are not sensitive to uncertainties associated
with the use of blank sky background datasets.

\section{Data Analysis}

\subsection{Cluster Mean Temperature and Fiducial Radius}
\label{sec:mean_kT}
The primary focus of this study is on cluster core properties, and we have
therefore chosen to devote our analysis to the main CCD chips for each of
the two detectors, i.e. chips 0--3 for ACIS-I and chip 7 for ACIS-S.  For
the spectral fitting, weighted response matrix files (RMFs) were generated
using the \CIAO\ task `mkacisrmf'. However, for some observations this was
not possible (as indicated in Section~\ref{sec:individual_clus}), owing to
a lack of calibration data suitable for mkacisrmf, and so the older task
`mkrmf' was used instead. In each case the correct gain file was used,
appropriate as described in the \CIAO\ documentation.

In a scaling study such as this, it is important to normalize observable
quantities appropriately, to provide a fair comparison between clusters of
different sizes. The key parameters of interest are mean gas temperature
and characteristic radius, and we outline below a simple scheme in which
both these quantities are determined self-consistently. Our fiducial radius
corresponds to an overdensity of 500 with respect to the critical density
of the Universe. This radius, \rfiveh, is commonly used for scaling studies
and corresponds to roughly 2/3 of \rtwoh\ \citep{san03b}.

We defer a full mass profile analysis and direct determination of 
overdensity radii to a later paper, and adopt in this work a simple, 
empirically calibrated proxy for the total mass within \rfiveh, based on 
the mean temperature, derived from the \MT\ relation of 
\citet{finoguenov01}. This has the advantage of permitting a direct 
comparison with other observations where poorer data quality prevents a 
full mass analysis being performed. For a cluster of redshift, $z$, the 
radius is given by \citep{willis05}
\begin{equation}
\rfiveh = 
\frac{391\times \Tbar^{0.63}}{E(z)}\,\, \mathrm{kpc} \label{eqn:rfiveh} 
\end{equation} where, \begin{equation}
 E(z) = (1+z) \sqrt{1+(z\,\,\omegam)+\frac{\omegal}{(1+z)^2}-\omegal}\,\,.
\end{equation}

In calibrating the temperature-radius relation using observed clusters, we
avoid the bias inherent in equivalent relations derived from numerical
simulations, which tend to overestimate the virial radii of the coolest
haloes \citep{san03}. In using the above expression for \rfiveh\ we have
not allowed for the effects of evolution in the overdensity factor
(i.e. the `500') with redshift. However, for such a low redshift sample,
the impact of this evolution is negligible.

The mean cluster temperature, \Tbar, is measured from the spectrum
extracted within 0.1--0.2 \rfiveh, thus excluding the central region where
the effects of strong gas cooling can contaminate the X-ray emission. The
combination of \Tbar\ and \rfiveh\ are determined iteratively, by
extracting an initial spectrum, finding the best-fit temperature, then
using this to estimate \rfiveh, and repeating the process until convergence
is achieved.

%
%
\begin{table*}
\begin{tabular}{l*{7}{c}}
\hline
Name & Mean kT & Core kT$^{a}$ & kT Ratio$^{b}$ & $R_{500}$ & CC significance$^{c}$ & Bautz-Morgan Type & Notes$^{d}$ \\
     & (keV)   & (keV)         &                &  (kpc)  &                       &                   &       \\
\hline\hline\\[-2ex] 
  NGC 5044 & 1.19 $ ^{+0.03}_{-0.02} $ & $ 0.88^{+0.00}_{-0.00} $ & $ 1.35^{+0.03}_{-0.03} $ & $ 437^{+6}_{-6} $ & 11.7* & -- & C, F \\
   Abell 262 & 2.34 $ ^{+0.06}_{-0.06} $ & $ 1.82^{+0.02}_{-0.02} $ & $ 1.28^{+0.04}_{-0.04} $ & $ 667^{+11}_{-11} $ & 7.9* & III & C \\
  Abell 4038 & 3.02 $ ^{+0.17}_{-0.17} $ & $ 3.26^{+0.14}_{-0.13} $ & $ 0.93^{+0.07}_{-0.07} $ & $ 785^{+28}_{-28} $ & -1.1 & III & -- \\
  Abell 1060 & 3.19 $ ^{+0.07}_{-0.07} $ & $ 3.40^{+0.04}_{-0.05} $ & $ 0.94^{+0.02}_{-0.02} $ & $ 811^{+12}_{-12} $ & -2.6 & III & -- \\
  Abell 1367 & 3.28 $ ^{+0.11}_{-0.10} $ & $ 3.32^{+0.10}_{-0.09} $ & $ 0.99^{+0.04}_{-0.04} $ & $ 826^{+17}_{-16} $ & -0.3 & II-III & M, R \\
 2A0335+096 & 3.47 $ ^{+0.14}_{-0.12} $ & $ 2.34^{+0.03}_{-0.03} $ & $ 1.48^{+0.06}_{-0.06} $ & $ 857^{+22}_{-18} $ & 8.2* & -- & C, F, M \\
  Abell 2147 & 4.17 $ ^{+0.14}_{-0.14} $ & $ 4.82^{+0.37}_{-0.19} $ & $ 0.87^{+0.06}_{-0.06} $ & $ 961^{+20}_{-20} $ & -2.3 & III & M \\
  Abell 2199 & 4.36 $ ^{+0.13}_{-0.11} $ & $ 3.93^{+0.04}_{-0.04} $ & $ 1.11^{+0.03}_{-0.03} $ & $ 989^{+18}_{-15} $ & 3.4* & I & C \\
   Abell 496 & 5.91 $ ^{+0.58}_{-0.56} $ & $ 3.58^{+0.11}_{-0.11} $ & $ 1.65^{+0.17}_{-0.17} $ & $ 1197^{+75}_{-70} $ & 3.9* & I & F \\
  Abell 2256 & 6.08 $ ^{+0.30}_{-0.31} $ & $ 5.64^{+0.45}_{-0.50} $ & $ 1.08^{+0.11}_{-0.11} $ & $ 1220^{+38}_{-38} $ & 0.7 & II-III & F, M, R \\
  Abell 1795 & 6.30 $ ^{+0.17}_{-0.17} $ & $ 4.88^{+0.06}_{-0.06} $ & $ 1.29^{+0.04}_{-0.04} $ & $ 1246^{+21}_{-21} $ & 7.5* & I & C, F \\
  Abell 3558 & 6.64 $ ^{+0.34}_{-0.51} $ & $ 6.04^{+0.32}_{-0.31} $ & $ 1.10^{+0.09}_{-0.09} $ & $ 1288^{+42}_{-62} $ & 1.1 & I & -- \\
    Abell 85 & 6.65 $ ^{+0.14}_{-0.14} $ & $ 4.80^{+0.06}_{-0.06} $ & $ 1.38^{+0.03}_{-0.03} $ & $ 1290^{+17}_{-17} $ & 11.4* & I & F, M, R \\
  Abell 3571 & 6.71 $ ^{+0.15}_{-0.42} $ & $ 7.85^{+0.17}_{-0.17} $ & $ 0.85^{+0.04}_{-0.04} $ & $ 1297^{+19}_{-51} $ & -3.6 & I & M \\
  Abell 3667 & 7.39 $ ^{+0.27}_{-0.27} $ & $ 6.76^{+0.18}_{-0.42} $ & $ 1.09^{+0.06}_{-0.06} $ & $ 1378^{+32}_{-32} $ & 1.5 & I-II & F, M, R \\
   Abell 478 & 7.97 $ ^{+0.18}_{-0.18} $ & $ 6.00^{+0.07}_{-0.07} $ & $ 1.33^{+0.03}_{-0.03} $ & $ 1446^{+21}_{-21} $ & 9.6* & -- & C,F \\
   Abell 401 & 8.35 $ ^{+0.43}_{-0.61} $ & $ 8.21^{+0.50}_{-0.63} $ & $ 1.02^{+0.09}_{-0.09} $ & $ 1488^{+48}_{-68} $ & 0.2 & I & M \\
  Abell 2029 & 9.11 $ ^{+0.20}_{-0.20} $ & $ 7.18^{+0.07}_{-0.07} $ & $ 1.27^{+0.03}_{-0.03} $ & $ 1572^{+21}_{-21} $ & 8.9* & I & F \\
  Abell 2142 & 9.45 $ ^{+0.65}_{-0.65} $ & $ 7.81^{+0.33}_{-0.33} $ & $ 1.21^{+0.10}_{-0.10} $ & $ 1610^{+71}_{-69} $ & 2.1 & II & F, M \\
  Abell 3266 & 9.86 $ ^{+0.53}_{-0.54} $ & $ 9.48^{+0.37}_{-0.37} $ & $ 1.04^{+0.07}_{-0.07} $ & $ 1653^{+57}_{-57} $ & 0.6 & I-II & M \\
\hline
\end{tabular}
\caption{
 Detailed properties of the sample, listed in order of increasing
 temperature. $^{a}$The core temperature measured within 0.1 \rfiveh;
 $^{b}$Ratio of mean to core temperature; $^{c}$Number of sigma difference
 from unity (Cool-core clusters are marked with ``*''). $^{d}$Clusters with
 detectable cavities in the ICM are marked with ``C'', cold front clusters
 are marked with ``F'', probable merger clusters are marked with ``M'', clusters
 with radio relics are marked with ``R''. All errors are 68\% confidence.  }
\label{tab:detailed}
\end{table*}

\begin{figure}
\includegraphics[width=8cm]{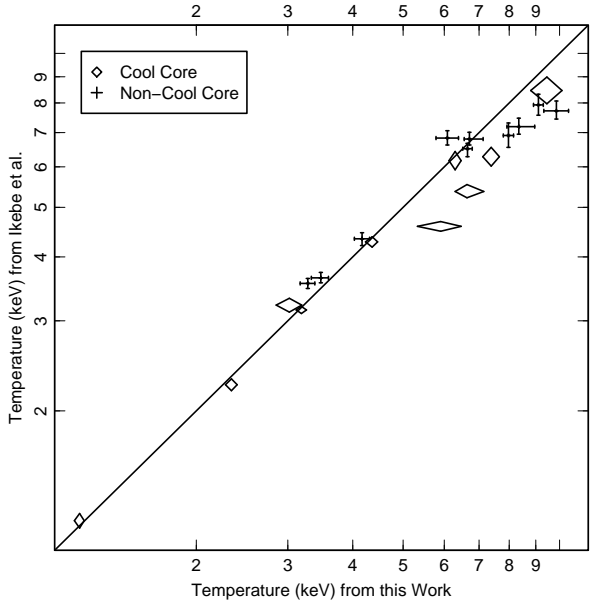}
\caption{ \label{fig:compare_kT}
 A comparison of the mean temperatures from this work and those of 
 \citet{ikebe02}, showing the line of equality.
}
\end{figure}

Fig.~\ref{fig:compare_kT} shows a comparison of the mean temperatures
derived with those from the original \ASCA\ analysis of
\citet{ikebe02}, who used a 2 temperature fit to correct for the 
effects of any cool core. There is good agreement below $\sim$5 keV, but
above this our temperatures are systematically hotter in almost every case,
for both CC and non-CC clusters. However, this behaviour is most likely the
result of differences in the regions from which spectra were extracted in
the two analyses. The most massive clusters have large angular sizes, which
are significantly larger than the \Chandra, but not \ASCA,
field-of-view. Since the temperature almost always declines beyond any cool
core \citep[\eg][]{mar98,vikhlinin05,piffaretti05}, this would lead to a
higher mean temperature measured with \Chandra.

\subsection{Cool Core Clusters}
\label{sec:CC}
Having determined the cluster mean temperature, unbiased by the effects of
central gas cooling, the spectrum of the core region ($<$0.1 \rfiveh) was
extracted separately and fitted as before to yield a core temperature (see
Table~\ref{tab:detailed}). A comparison between this core temperature and
the mean (core-excluded) temperature can then be used to quantify the
influence of central cooling. By considering the ratio of the mean to the
core temperature as a discriminator, we define cool core (CC) clusters as
those systems for which this ratio exceeds unity at greater than 3$\sigma$
significance. This provides a clean separation of the sample into 9 CC and
11 non-CC clusters, with a mean ratio and standard deviation of 1.35/0.15
and 1.0/0.11, respectively. 

For comparison, in a recent \Chandra\ study \citet{bauer05} find that at
least 55 per cent of their sample of 38 X-ray luminous clusters show signs
of mild cooling, with 34 per cent displaying evidence of strong
cooling. According to the definition used by \citeauthor{bauer05}, all the
clusters in our sample could be classified as having a cool core (cooling
time, \tcool $<$ 10 Gyr). However, their sample is more distant
($z\sim$ 0.15--0.4), and they are correspondingly less able to resolve the
inner parts of the ICM where the gas cooling time is lowest and possibly
falls below 10 Gyr in many cases. The older study of \citet{peres98}
reported a cool-core fraction of 70 per cent, based on lower-resolution
\ROSAT\ observations of a complete sample of the 55 brightest clusters in
the sky in the 2--10 keV band. However, their definition of a cool core is 
different again, requiring that the upper limit to the central \tcool\ be
less than the assumed age for the cluster.

In any case, both \citet{bauer05} and \citet{peres98} base their definition
of a cool-core on the gas cooling time which, as they demonstrate, is
clearly capable of reaching low values ($\sim$few Gyr) in most --- perhaps
all --- clusters (see also Section~\ref{sec:tcool}). By casting our
definition in terms of a significant temperature decrease in the inner
0.1\rfiveh, we are able to identify clusters where radiative cooling has
demonstrably impacted the gas properties in a substantial way, beyond
merely forming at least some gas with a low cooling time.

\subsection{Spectral Profiles and Deprojection Analysis}
\label{ssec:spec_prof}
In order to study the spatial variation of gas temperature, a projected
temperature profile was obtained for each cluster. Spectra were extracted
in a series of concentric annuli, centred on the peak of the X-ray
emission. The radial bins were chosen to enclose a fixed number of net
cluster counts between 1000 and 3000, depending on the quality of the
observation. Each spectrum was fitted with an absorbed APEC model, as
above, to yield the best-fit temperature. A characteristic radius was
assigned to each annulus using the emission-weighted approximation of
\citet{mcl99},
\begin{equation}
r = \left[ 0.5\left( \rmsub{r}{out}^{3/2} + \rmsub{r}{in}^{3/2} \right)
\right]^ {2/3} .
\end{equation}

To derive estimates of the 3 dimensional gas temperature, we used the
\XSPEC\ PROJCT model to deproject the spectral profiles under the
assumption of spherical geometry. The deprojection is quite slow and
susceptible to strong noise fluctuations, so we used a smaller number of
coarser annular bins --- between 10 and 20, according to data quality. To
stabilize the fitting, the absorbing column and gas metallicity we fixed at
values obtained by fitting each annulus separately prior to the
deprojection. For some clusters, it was necessary to freeze the absorbing
column at the galactic HI value (as detailed in Table~\ref{tab:sample}),
since unfeasibly low values were otherwise obtained in many of the annular
bins. However, we verified that in those bins where the absorbing column
was able to be fitted, the optimum values were fully consistent with the HI
inferred measurement. Moreover, we found no indication of any radial trend
in absorbing column in these cases, which might point to problems with the
calibration. Similarly, in a number of cases the deprojected temperature
had to be fixed at its projected value, to produce a stable fit (denoted by
a `T' in the rightmost column of Table~\ref{tab:sample}). However, these
are all non-CC clusters, with approximately isothermal temperature
profiles, where the smoothing effects of projection are minimal, thus any
bias introduced by this approximation is likely to be small.

\section{Notes on Individual Clusters}
\label{sec:individual_clus}
In this section, we provide further information about each of the clusters
in the sample, highlighting key aspects of the analysis specific to
different datasets.

\subsection{NGC 5044}
This is the only galaxy group in the sample, since a selection based on 
flux biases towards more massive clusters. It is a well-studied object, 
which hosts a cold front and cavities seen in X-ray emission \citep{buote03}.

\subsection{Abell 262}
A known cavity cluster \citep{blanton04}. The optical properties of this
cluster are unusual, and its luminosity function and colour-magnitude
relation shown signs of contamination from a large number of lower mass
galaxies, possibly associated with a nearby supercluster (W. Barkhouse, 
private communication).

\subsection{Abell 4038}
It was necessary to use the projected temperature profile in place of the
deprojected one for this cluster, in order to stabilize the fitting and
thereby avoid strong fluctuation between adjacent bins.

\subsection{Abell 1060}
Optical observations indicate that A1060 has a dynamically perturbed
condensed core \citep{girardi97}. However, no obvious merger activity is
evident from the X-ray emission, so this cluster has not been classified as
a merger candidate in Table~\ref{tab:sample}. The projected $T(r)$ data
were used for the deprojection, to avoid strong fluctuations in the
recovered profile.

\subsection{Abell 1367}
This is a complex structured cluster, which is clearly undergoing multiple
merging, as seen in both optical \citep[e.g.][]{cortese04} and X-rays
\citep{sun02}. The projected $T(r)$ data were used for the deprojection, to 
avoid strong fluctuations in the recovered profile.

\subsection{2A0335+096}
This is known to be a cold front cluster \citep{mazzotta03}, which also
possesses a radio-lobe cavity \citep{birzan04}. A recent deep \XMM\ analysis
of this system indicates that it is also likely to be undergoing a merger
with a subcluster \citep{werner06}.

\subsection{Abell 2147}
This cluster is composed of several clumps, with evidence of luminosity
segregation in the galaxies \citep{lugger89}; it has thus been labelled as
a merger candidate in Table~\ref{tab:sample}. It is interesting to note
that the brightest cluster galaxy for A2147 is not located at the centre of
the cluster potential (and X-ray peak) \citep{lugger89}.

It was necessary to fix the absorbing column at the galactic HI value,
since unfeasibly low values were obtained when it was left free to
vary. The projected $T(r)$ data were used for the deprojection, to avoid
strong fluctuations in the recovered profile.

\subsection{Abell 2199}
This is a known cavity cluster \citep{johnstone02}.

\subsection{Abell 496}
This cluster hosts a prominent cold front \citep{dupke03}.

\subsection{Abell 2256}
This is a probable cold front cluster in the early stages of merging
\citep{sun02b}. Our annular spectral bins were centred on the main (East) 
cluster peak, referred to as `P1' in the analysis of \citet{sun02b}. 

It was necessary to use the projected $T(r)$ for the deprojected $T(r)$, to
avoid strong fluctuations in the deprojected profile, arising from a
complicated `S' shaped profile. The absorbing column was also fixed at
the galactic HI value for all bins, to stabilize the fitting; there is no
evidence from the global spectrum of any significant difference between the
fitted column and the HI value. It was necessary to use the older \CIAO\ task
`mkrmf' to generate spectral responses for this dataset.

\subsection{Abell 1795}
This is a cold front cluster \citep{markevitch01}.

\subsection{Abell 3558}
It was necessary to use the projected $T(r)$ for the deprojected $T(r)$ for
this cluster.

\subsection{Abell 85}
A well-known subclump cluster, which also has a cold front and is a 
probable merger candidate \citep{kempner02}. The prominent Southern subclump 
and also the Western subclump were both excluded from the analysis. 

\subsection{Abell 3571}
This cluster is probably in the late stages of merging \citep{venturi02}
and is located in the Shapley supercluster.  It was necessary to use the
projected $T(r)$ for the deprojected $T(r)$.

\subsection{Abell 3667}
This is both a cold front and merging cluster \citep{vikhlinin01}. It was
necessary to use the projected $T(r)$ for the deprojected $T(r)$.

\subsection{Abell 478}
This well known cool core cluster hosts a cold front \citep{markevitch03} and
possesses prominent X-ray cavities with coincident radio lobes \citep{sun03}.

\subsection{Abell 401}
This is a probable merger remnant \citep{sakelliou04}. The older \CIAO\ task
`mkrmf' was used to created a spectral response matrix, owing to lack of
calibration data for the newer task `mkacisrmf'.  It was necessary to use
the projected $T(r)$ for the deprojected $T(r)$.

\subsection{Abell 2029}
Another well studied cool core cluster which possesses a cold front
\citep{markevitch03}.

\subsection{Abell 2142}
This is the archetypal `cold-front' cluster, which likely consists of the
largely intact core of a poor cluster enclosed within a halo of much hotter
gas associated with a larger cluster (Markevitch et al. 2000).
This fact accounts
for its unusual temperature profile, which is atypical of non-CC clusters 
(see section~\ref{ssec:T(r)}).

The two observations of Abell~2142 (obsid 1196 \& 1296) were amongst the
very first made by \Chandra, and were taken when the CCD temperature was
only -100C. Since the standard \CIAO\ calibration does not cover this CCD
temperature, these datasets were analysed according to the calibration for
a CCD temperature of $-110$C. In addition, the corresponding Markevitch
period A blank sky background datasets are truncated at 10 keV, so the
background renormalization was performed in the range 8--10 keV. The
analysis presented here is based on the 1196 dataset, but to verify our
results we also analysed the 1296 observation, and found very good
agreement between the density and temperature profiles. As for A401, it was
necessary to use the older \CIAO\ task `mkrmf' to generate spectral
responses for this dataset.

\subsection{Abell 3266}
This is a well-known merging cluster \citep{henriksen02}. It was necessary
to use the projected $T(r)$ for the deprojected $T(r)$.

\section{Results}

\subsection{Temperature Profiles}
\label{ssec:T(r)}
\begin{figure*}
\centering
\includegraphics[width=16.6cm]{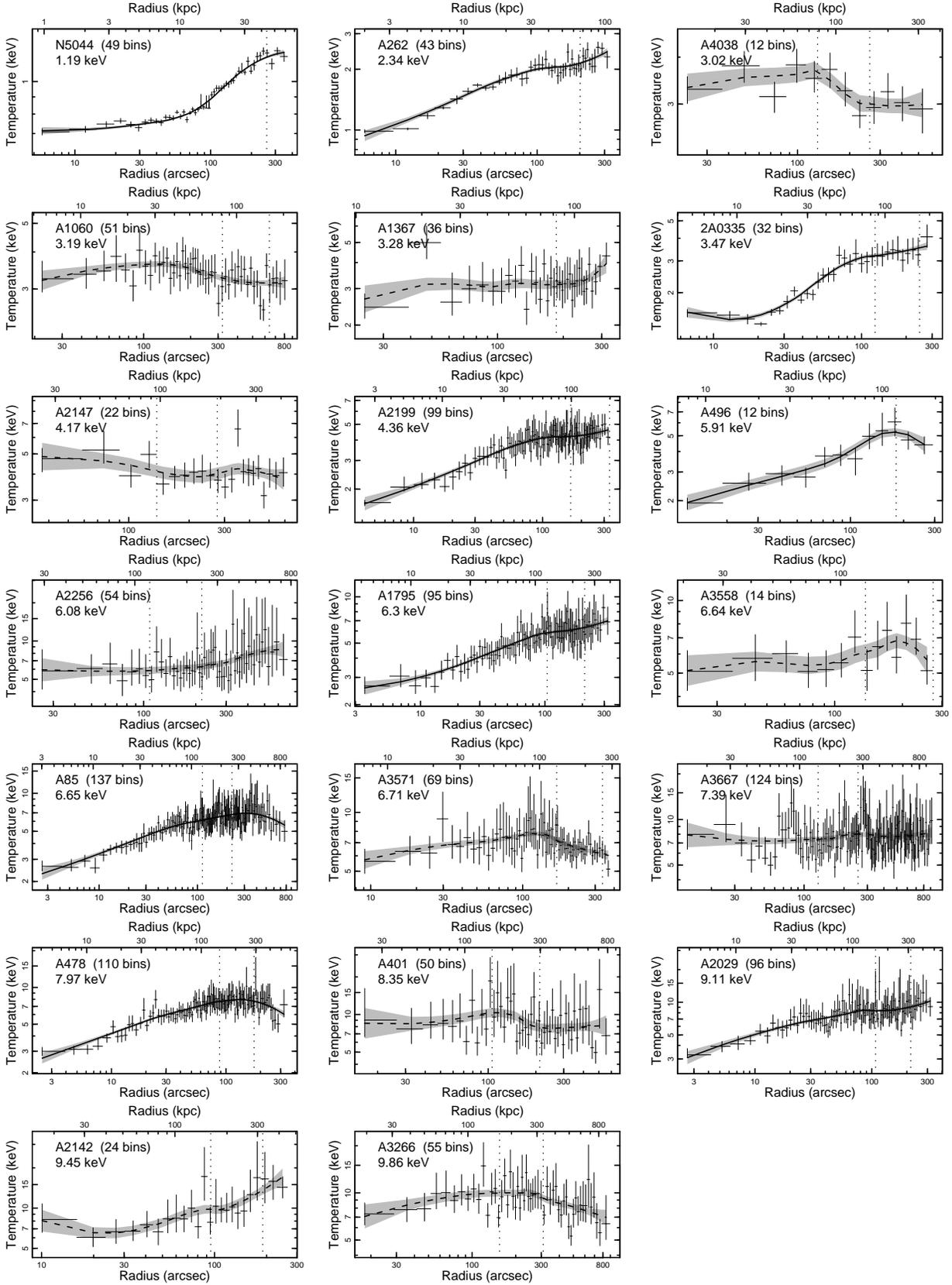}
\caption{ \label{fig:mosaic_T(r)}
 Projected temperature profiles and 1$\sigma$ error envelopes for each
 system, ordered by increasing mean temperature from top left to bottom
 right. The curves represent a locally-weighted fit to the data points (in
 log-log space). Cool-core clusters are denoted by solid lines and non
 cool-core clusters are plotted with dashed lines. Vertical dotted lines
 bracket the region 0.1--0.2 \rfiveh, within which the mean temperature is
 measured. The mean temperature for each cluster is indicated, as is the
 number of annular bins used to construct the curves.}
\end{figure*}

\begin{figure*}
\centering
\includegraphics[width=18cm]{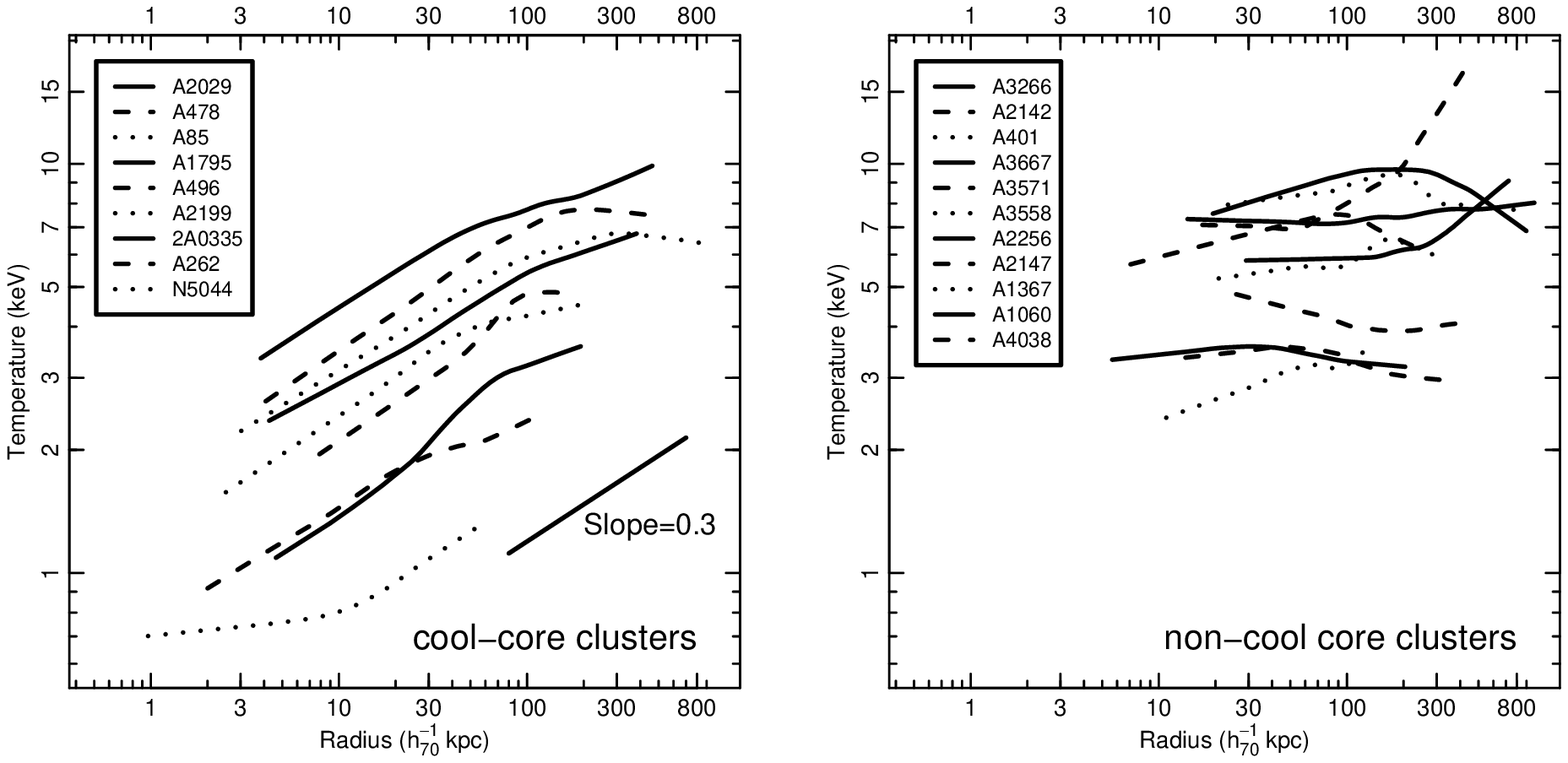}
\caption{ \label{fig:kpc_T(r)}
  Projected, finely-binned temperature profiles for the sample, split by
  cool core type and using an arbitrary line style to uniquely identify
  each system: the cluster name entries in the legends are ordered by
  temperature. Both plots are scaled identically in temperature and
  radius. Each curve represents a locally-weighted fit to the data points
  (in log-log space), to suppress scatter (see text for details).  }
\end{figure*}

The temperature profiles from the projected annular analysis are plotted
separately for each cluster in Fig.~\ref{fig:mosaic_T(r)}, with vertical
dotted lines indicating the region 0.1--0.2 \rfiveh\ used to calculate the
mean temperature. To clarify the underlying trend in each profile, the raw
data points have been fitted with a locally weighted regression in log-log
space, using a similar method to that outlined in \citet{san05}.  This
technique effectively smooths the data using a quadratic function which is
moved along the set of points to build up a curve, in an analogous fashion
to the way a moving average is computed for a time series. The algorithm
used is implemented in the `loess' function in version 2.3 of the \Rproject\
statistical environment package\footnote{http://www.r-project.org}
\citep{Rcite}, which also provides an estimate of the 1$\sigma$ error on 
the regression, based on the scatter about the curve. To compare profiles
across the sample, the projected temperature are plotted as a function of
physical radius (in kpc) for CC and non-CC clusters in separate panels in
Fig.~\ref{fig:kpc_T(r)}. Once again the curves represent a locally-weighted
regression, this time using the \Rproject\ `lowess' function \citep[as used
in][]{san05}, which provides a larger degree of smoothing. The curves have
been assigned an arbitrary line style to uniquely identify each system.

The difference between CC and non-CC clusters is very clear from
Fig.~\ref{fig:kpc_T(r)}. Within the cool core, there is a remarkable
consistency in shape of the temperature profile, which follow a power law
with a logarithmic slope of roughly 0.3.  In most clusters this behaviour
appears to extend down to a few kiloparsecs --- well within the central
galaxy. The obvious exception to this is the coolest system, the NGC~5044
group, which exhibits a sharp break at $\sim$10 kpc, as the logarithmic
slope flattens within the central galaxy. As this is the only galaxy group
in the sample, this result may point to a different scaling of hot gas at
small mass scales. However, this system does display evidence of multiphase
gas and morphological disturbances, in the form of X-ray cavities from AGN
activity as well as a cold front \citep{buote03}. It is therefore possible
that this flattening in $T(r)$ is an atypical feature. However, we
emphasize that the possibility of a flattening of $T(r)$ in the innermost
core of clusters \citep[as recently claimed by][for example]{donahue06} is
not ruled out by our data, since the `lowess' regression tends to mask such
behaviour, by down-weighting any deviations of the innermost few points
from the general trend. This disadvantage is compensated for by the clarity
with which the local regression captures the underlying trend in $T(r)$, to
highlight the scaling between different clusters.

\begin{figure}
\includegraphics[width=8cm]{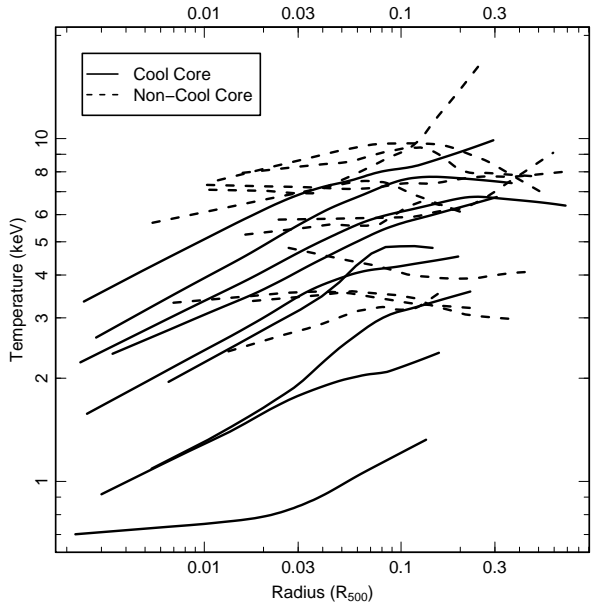}
\caption{ \label{fig:2d_T(r)}
  Projected, finely-binned temperature profiles for each of the clusters,
  scaled to \rfiveh\ and labelled according to the presence of a cool
  core. Each curve represents a locally-weighted fit to the data points,
  to suppress scatter (see text for details).  }
\end{figure}

Fig.~\ref{fig:2d_T(r)} shows the variation in projected temperature with
scaled radius. The size of the cool core region appears to be reasonably
well approximated by 0.1\rfiveh --- the radius of the core exclusion region
used in calculating the mean cluster temperature
(Section~\ref{sec:mean_kT}). However, any turn-over in the temperature
profile at larger radius is difficult to detect, given the limited
field-of-view of the main \Chandra\ detectors. This is compounded by the
fact that all but one of the CC cluster observations analysed here used the
ACIS-S detector, which covers 1/4 of the area of ACIS-I. A further
systematic difference is evident at smaller radii, where the profiles in
Fig.~\ref{fig:kpc_T(r)} extend further in for CC clusters, as a consequence
of the bright cool core enabling finer annular binning.

\begin{figure}
\includegraphics[width=8cm]{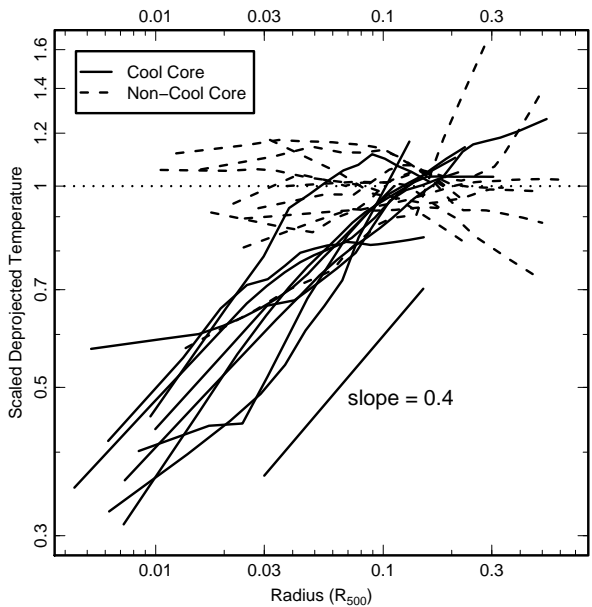}
\caption{ \label{fig:scT(r)}
  Deprojected temperature profiles for CC and non-CC clusters, scaled
  by \Tbar\ and plotted as a function of scaled radius. Each curve
  represents a locally-weighted fit to the data points, to suppress scatter
  (see text for details). }
\end{figure}

The individual deprojected \emph{scaled} temperature profiles are plotted
in Fig.~\ref{fig:scT(r)}, as a function of scaled radius. Here each profile
has been normalized by the mean temperature of the cluster, to provide a
direct comparison between systems of different mass. It is clear that there
is a tight spread between the cool core clusters, which hold to an
approximately power-law form, with a logarithmic slope of $\sim$0.4
(plotted as a solid line with arbitrary normalization). Conversely, the
non cool-core clusters scatter around the locus of isothermality (dotted 
line), showing no obvious systematic variation with radius.

\begin{figure}
\includegraphics[width=8cm]{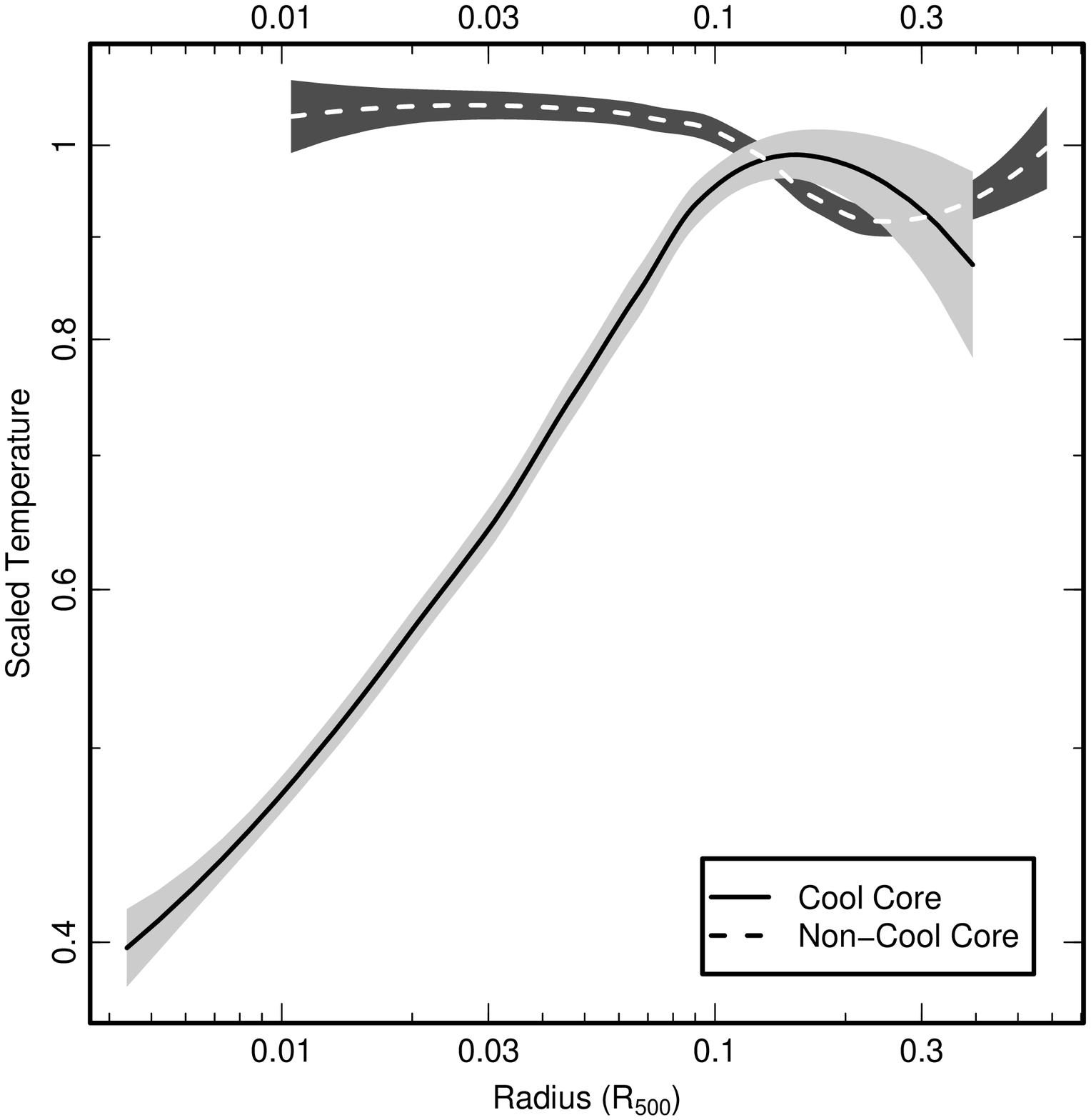}
\caption{ \label{fig:mean_scT(r)}
  Mean deprojected temperature profiles for CC and non-CC clusters, scaled
  by \Tbar\ and plotted as a function of scaled radius. Each curve
  represents a locally-weighted fit to the data points, to suppress scatter
  (see text for details). The shaded regions represent the 1$\sigma$
  confidence region of the fit. }
\end{figure}

Motivated by the uniformity in temperature profiles displayed by the CC
clusters, we have combined all the scaled data points for these systems and
subjected them to a locally-weighted regression, to determine a
characteristic average temperature profile. Each cluster's deprojected
temperature profile was scaled by the mean cluster temperature (\Tbar) and
corresponding \rfiveh. The regression was performed in log-log space, using
the \Rproject\ task `loess', in order to provide an estimate of the error on
the regression, from the scatter about the line. Each point was weighted by
its inverse variance (computed using the mean asymmetric error), to improve
the rejection of outliers. The same procedure was applied to the non-CC
clusters, to provide a comparison. The corresponding profiles and 1$\sigma$
error envelopes for the CC and non-CC clusters are plotted in
Fig.~\ref{fig:mean_scT(r)}. The outermost point has been excluded from the
mean CC and non-CC profiles plotted, in order to minimize any bias that
this single point can cause at such large radius.

The behaviour of the profiles at large radii is less certain, as indicated
by the widening of the regression confidence region, resulting from the
diminishing number of points. However, there is clearly a preference for a
peak at $\sim$0.15 \rfiveh, followed by a decline, which is broadly
consistent with previous analyses
\citep[e.g.][]{vikhlinin05,piffaretti05,zhang06}.

\subsection{Cooling Time}
\label{sec:tcool}
\begin{figure}
\includegraphics[width=8cm]{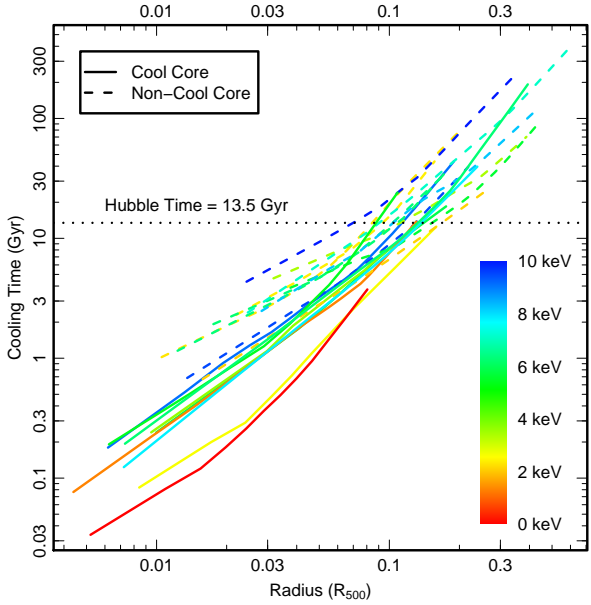}
\caption{ \label{fig:t_cool(r)}
  Gas cooling time profiles for each of the clusters, scaled to \rfiveh,
  labelled according to the presence of a cool core, and coloured according
  to the cluster mean temperature. Each curve represents a locally-weighted
  fit to the data points, to suppress scatter (see text for details).  }
\end{figure}

A simple estimate of the susceptibility of gas to radiative cooling can be
obtained by considering the timescale over which the gas can continue to
lose energy at its current rate. Thus defined, the cooling time (in
seconds) of a parcel of gas with volume, $V$ (in cm$^3$), and luminosity,
\LX\ (in erg s$^{-1}$), is given by
\begin{equation}
 \tcool = 1.602\times10^{-9} \times \frac{3 kT \rho V \rmsub{\mu}{e}}{2 \mu \LX} ,
\label{eqn:t_cool}
\end{equation}
where $kT$ is the \emph{deprojected} gas temperature (in keV), $\rho$ is
the electron number density (in cm$^{-3}$) and the constants \rmsub{\mu}{e}
and $\mu$ are the mean mass per electron (1.167) and the mean molecular
weight (0.593) of the gas. Profiles of cooling time as a function of scaled
radius are plotted in Fig.~\ref{fig:t_cool(r)}, colour coded by the mean
cluster temperature. As before the curves represent a locally-weighted
regression to the data points. Note that the cooling time in the outermost
annular bin is unknown (unlike the temperature), since the assumed volume
is not well defined, which means that the gas density cannot be determined.

\begin{figure}
\includegraphics[width=8cm]{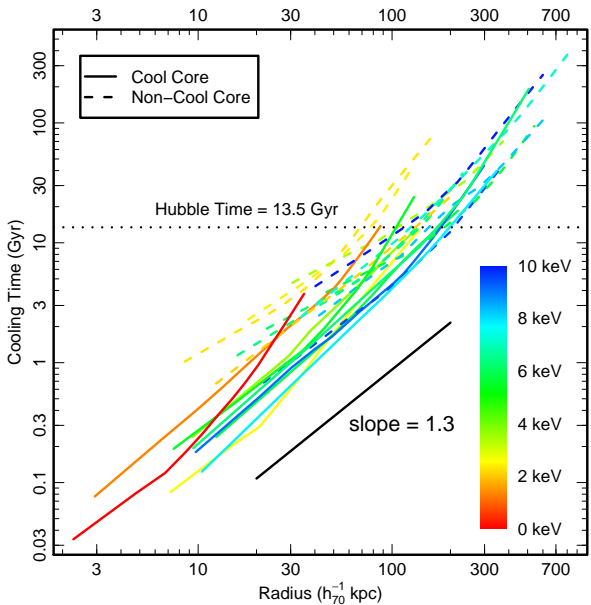}
\caption{ \label{fig:t_cool(r_kpc)}
  Same as Fig.~\ref{fig:t_cool(r)}, but plotted as a function of radius in
  kpc.}
\end{figure}

For purely bremsstrahlung X-ray emission, $\tcool \propto
\sqrt{kT}$. However, the contribution of line emission becomes significant
at lower temperatures, which weakens this trend. Therefore this simple
scaling has not been applied to Fig.~\ref{fig:t_cool(r)}, and the profiles
are thus largely ordered by mean cluster temperature. Nonetheless, despite
this spread in temperature, there is a clear consistency between all the
profiles, with much greater similarity between CC and non-CC clusters than
is seen in the gas temperature. Not surprisingly, the cool core systems
have consistently lower cooling times at all radii, but they also exhibit a
slightly steeper logarithmic slope than those of the non-CC clusters. While
it is clear that CC clusters reach much lower values in the core, it can be
seen that within $\sim$0.1\rfiveh, most non-CC clusters have rather low
cooling times, falling below a Hubble time in all cases. 

It can be seen in Fig.~\ref{fig:t_cool(r)} that the locus corresponding to
the Hubble time intersects the profiles around 0.1\rfiveh. This is the same
radius used to excise any cooling-dominated flux from the mean cluster
temperature estimates (Section~\ref{sec:mean_kT}), and also approximately
corresponds to the point where the CC cluster temperature profiles reach a
maximum (Fig.~\ref{fig:mean_scT(r)}). It is apparent that even in non-CC
clusters, the cooling time of gas within this radius can fall well below a
Hubble time. However, the non-CC gas cooling time seems to reach a minimum
level of around 1 Gyr, whereas all the CC profiles fall well below this
value. To some extent this behaviour can be explained by the fact that cool
core clusters permit a much finer scale spectral binning, and so their gas
properties can be measured to smaller radii, where \tcool\ is naturally
lower. Nevertheless, non-CC clusters have noticeably shallower cooling time
profiles at all radii.

An alternative perspective is obtained by studying cooling time as a
function of physical rather than scaled radius, as plotted in
Fig.~\ref{fig:t_cool(r_kpc)}. Here the scatter between the profiles is
actually reduced, particularly beyond $\sim$100 kpc. Moreover, the two
largest outliers in Fig.~\ref{fig:t_cool(r)} (the coolest two systems) lie
well within the main trend, albeit with NGC~5044 showing a significantly
steeper outer slope. The implication of this is that the cooling time of
gas in virialized systems (either with or without a cool core) obeys a
universal form across more than two orders of magnitude in radius, with
relatively small scatter (a factor of $\sim$3), as found previously by
\citet{voigt04} and \citet{bauer05}. This suggests a prominent
role for radiative cooling in governing cluster properties. It is not clear
exactly why this should be so, but it represents an interesting challenge
to theoretical models of cluster properties to account for this behaviour.

Within $\sim$100 kpc \tcool\ scales roughly as $r^{1.3}$, as indicated by
the solid black line, which agrees with the findings of \citet{voigt04},
based on a sample of 16 clusters with strong cool cores. A similar result
was obtained by \citet{bauer05}, from a sample of 38 X-ray luminous
clusters at moderate redshift. Furthermore, the apparent steepening of the
profiles at larger radius is also consistent with both these previous
studies.

\section{Discussion}
\label{sec:discuss}
It is clear that clusters can be subdivided into two distinct categories
according to the presence or absence of a cool core 
\citep[e.g.][]{peres98,bauer05}. How then can we explain the mixture of 
these types in the cluster population? Judging from our cooling time
results, gas in the cores of all clusters falls well below a Hubble time
(down to a least $\sim$1 Gyr). This demonstrates that the intracluster
medium is certainly capable of developing a cool core in the inner regions;
that it does not always do so implies the influence of some counteracting
process(es).

Given the predominantly isothermal temperature profiles seen amongst the
non-cool core clusters, it is natural to examine the role of cluster
merging in erasing cool cores, which would tend to mix gas and smooth out
thermal gradients \citep[e.g.][]{fabian84,edge92,buote96}. The rightmost
column of Table~\ref{tab:detailed} indicates which clusters show signs of
merging activity, and a clear trend is apparent: 8 out of 10 `merger'
clusters are non-CC, leaving only 3 out of 11 non-CC clusters showing no
obvious signs of merging, namely Abell~1060, Abell~4048 and Abell~3558. Of
these, A1060 has a dynamically perturbed condensed core \citep{girardi97}
and may be a relatively recent merger cluster in which a cool core has not
yet been re-established \citep{yamasaki02}. The CC clusters with
indications of merging activity are Abell~85 and 2A0335+096. Of these, A85
clearly contains two subclusters and it appears that the cool core has
survived a merger with at least one of these objects \citep{kempner02}. A
similar event may well be occurring in 2A0335+096 \citep{werner06}, thus
demonstrating that not all mergers are capable of destroying an existing
cool core.

If merging is primarily responsible for erasing cool cores, then its impact
outside this region must be relatively minor. We note that there is no
evidence that the non-CC clusters are significantly hotter on average than
the CC ones, (e.g. Fig.~\ref{fig:z-kT}), based on the temperature measured
within 0.1--0.2 \rfiveh\ (Section~\ref{sec:mean_kT}); a comparison of the
temperatures of the CC and non-CC clusters, using the Kolmogorov-Smirnov
Test (implemented in the \Rproject\ function `ks.test'), results in a
$p$-value of 0.9195 in favour of the null hypothesis that the two sets of
data are drawn from the same continuous distribution. It can be seen that
the cooling time profiles of both types of cluster roughly converge beyond
$\sim$0.1\rfiveh\ (Fig~\ref{fig:t_cool(r)}). The head-on merger simulations
of \citet{gomez02} indicate that the residual signature of a cool core may
remain for up to 1--2 Gyr after a disruptive merger, before the onset of
turbulence heats the gas significantly. Thus it is possible for a cool core
cluster to have recently experienced a significant disruption, which could
account for the presence of cool cores and merger signatures in both A85
and 2A0335+096.

Within the core, the apparent `floor' value of $\sim$1 Gyr seen in the
cooling time of non-CC clusters could represent the time taken to re-form a
cool core. If mergers are primarily responsible for erasing cool cores,
then this timescale can be linked to the characteristic timescale between
such events. Specifically, the fraction of non-CC clusters would correspond
to the fraction of the time between mergers that is required to
re-establish a cool core. Therefore, if the $\sim$1 Gyr `floor' value
represents the time taken to re-form a cool core and roughly 50 per cent of
the sample possess a cool core, this implies a characteristic time of
$\sim$2 Gyr between mergers. By comparison, the simulations of
\citet{cohn05} indicate that, on average, clusters in their sample of 574
have experienced at least 4 large mergers since $z\sim2$, corresponding to
a lookback time of $\sim$10 Gyr. This implies a mean timescale of roughly
2.5 Gyr between mergers, which is consistent with the above figure.

Alternatively, cool cores could be disrupted by AGN-driven outflows
originating in the central galaxy, which is a plausible mechanism for
explaining their absence \citep[e.g.][]{mcnamara05,osu05a} and is capable
of mixing up gas \citep[e.g][]{osu05b}. However, all of the six clusters in
our sample with obvious evidence of AGN-associated cavities in the ICM
(marked with a `C' in the rightmost column of Table~\ref{tab:detailed})
possess cool cores, reflecting the well established trend for cool core
clusters to host central radio sources \citep{burns90}. On the other hand,
AGN may well play an important role in regulating unchecked cooling in
cluster cores, which is well established to be significantly suppressed
\citep[and references therein]{peterson06}. If AGN activity contributes 
towards offsetting cooling in addition to merging, then this would impact
the merger rate inferred above. Specifically, any contribution from AGN
heating would require correspondingly fewer mergers, thereby increasing the
timescale between such events. In that case, the value of $\sim$2 Gyr
deduced above would represent a lower limit to the time between major
mergers.

For purely bremsstrahlung X-ray emission the emissivity scales as
$\sqrt{kT} \rhogas^2$, which in turn implies $\tcool \propto
\sqrt{kT}$. Therefore, the apparent universality of the cooling time as a
function of physical radius (Fig.~\ref{fig:t_cool(r_kpc)}) indicates that
hotter clusters must have a higher gas density (approaching $\rhogas
\propto \sqrt{kT}$). However, a given physical radius corresponds to a 
smaller fraction of \rfiveh\ in hotter clusters, which therefore samples
gas of higher density, since \rhogas\ generally increases towards the
centre. Thus, even if gas at a given fraction of \rfiveh\ had the same
density in clusters of all masses, there would be some reduction in the
scatter in \tcool\ plotted against physical radius compared to
\rfiveh. Nevertheless, a trend for a lower gas density in groups compared
to clusters has previously been reported
\citep[e.g.][]{san03}. Furthermore, the generally good agreement found
between the baryon fraction in massive clusters and the the ratio of
$\rmsub{\Omega}{b}/\rmsub{\Omega}{m}$ \citep[e.g.][]{allen02,san03b},
demonstrates that it is the reduced \rhogas\ in less massive systems that
is anomalous.

If clusters and groups form self-similarly at any given epoch from material
of constant density then radiative cooling, or indeed non-gravitational
heating --- operating more effectively in groups --- could have acted to
deplete the gas in the centres of groups, thereby lowering the density of
the remaining material. However, such a process must have occurred
subsequent to collapse and virialization, and so we would expect to observe
evidence of systems in a intermediate state, which would register as
outliers in Fig.~\ref{fig:t_cool(r_kpc)}. The lack of significant outliers
suggests that the implied variation in gas density may already have
substantially been in place \emph{prior} to virialization, which implies
\emph{non}-self-similar accretion.

Such a scenario is consistent with models in which preheating takes places
predominantly in filaments, whose linear geometry causes the energy
injected to impact the density in preference to the temperature \citep[as
suggested by][]{pon03,voit03b}. Thus, the gas in the smaller filaments
feeding small haloes could be `puffed up' more than that in larger cluster
filaments, lowering the density of material accreted onto groups. However,
if this picture is correct, it points to an uncomfortable coincidence,
namely that preheating of gas prior to accretion and shock heating produces
virialized material of roughly constant cooling time at any given physical
radius. Such an apparent conspiracy might only be resolved if some
fundamental connection between preheating and cooling in haloes could be
established.

\subsection{Mergers, Radio Relics and Cold Fronts}
Many clusters are known to possess large-scale, diffuse radio sources which
are related to the intracluster medium rather than AGN
\citep{giovannini04}. Such sources emit synchrotron radiation, and are
referred to as radio relics. Of the clusters in this sample, four are
classified as radio relics, namely Abell~85 \citep{giovannini00}, Abell~1367
\citep{gavazzi78}, Abell~2256 \citep{rottgering94, clarke06} and Abell~3667 
\citep{rottgering97}. The non-CC cluster Abell~4038 also has a radio halo 
similar to a relic, but this is most likely the remnant of a radio galaxy
now located $\sim$18 kpc to the East of the relic
\citep{slee01}. Similarly, Abell~2142 has a small radio halo which may
originate from a single galaxy which was active in the past
\citep{giovannini00}. The generally accepted view is that genuine radio
relics (i.e. much larger scale than emission from a single radio galaxy)
are the product of shock waves arising from merging activity
\citep{ensslin98}, and this picture is borne out by the fact that all 
four of these clusters are merger candidates. However, A85 also possesses a
cool core, demonstrating that it is possible for a cool core to survive a
merger energetic enough to produce a radio relic.

A related aspect is the prevalence of cold fronts in the sample, which are
found in 7 of the CC clusters and 3 of the non-CC clusters. Of these 10
cold front clusters, 5 show signs of merging. Clearly cold fronts are much
more common in cool core clusters (7/9) than non-CC clusters (3/11), which
is consistent with the findings of \citet{markevitch03}, who reported that
roughly 70 per cent of the nearby cool core clusters in their \Chandra\
sample of 37 contained cold fronts. A similar fraction has been found more
recently from an \XMM\ analysis of 62 clusters \citep{ghizzardi05}.  In any
case, the presence of cold fronts is certainly not incompatible with the
existence of a cool core. For very relaxed clusters, like Abell~2029 and
Abell~1795, cold fronts may result from the infall of small subhaloes
\citep{ascasibar06}, rather than being caused by more substantial merging.

\subsection{Properties of the Central Galaxy}
\label{ssec:BM}
Most of the clusters in this sample possess a large central galaxy which is
coincident with the X-ray peak. The presence of this additional potential
well is likely to impact the cluster properties in the core, so here we
briefly explore the characteristics of the central galaxy. The Bautz-Morgan
(BM) classification scheme \citep{bau70} was devised to catalogue clusters
of galaxies according to the contrast between the central galaxy and the
other galaxies in the cluster \citep[see][for a review]{bah77}. There are
three main categories, which are defined as follows:
\begin{description}
\item Type I are clusters containing a centrally located cD galaxy
\item Type II are clusters where the brightest members are intermediate in
  appearance between cD galaxies and Virgo- or Coma-type normal giant
  ellipticals.
\item Type III are clusters containing no dominant galaxies.
\end{description}

Although the BM type is a subjective quantity, which is susceptible to bias
\citep{leir77}, it none the less provides a reasonable measure of the
evolutionary state of the cluster. BM classifications are available for 17
of the 20 clusters in the sample (the 3 unclassified systems are Abell 478,
NGC~5044 and 2A0335+096) and are listed in Table~\ref{tab:detailed}. A
total of 5 out of the 6 CC clusters with a BM classification contain type~I
central galaxies, compared to only 3 out of 11 non-CC clusters. Moreover,
the only CC cluster with a non-type~I classification is an anomalous
system: Abell~262 (type~III) exhibits a very unusual colour magnitude
relation and luminosity function, with an apparent large excess of fainter
galaxies, consistent with significant contamination from a background
supercluster (W. Barkhouse, Private Communication).

It is clear that BM type I clusters are more evolved than type II and III,
in the sense that they are unlikely to have experienced significant recent
disruption. The tendency for CC clusters to be BM type I and non-CC
clusters to be types II and III is therefore consistent with the hypothesis
that merger activity removes or otherwise prevents the formation of a cool
core. This behaviour is consistent with the findings of \citet{buote96},
who discovered a tendency for larger BM types to show evidence of global
morphological disturbance in the form of a larger power ratio. Similarly,
\citet{ledlow03} find that earlier BM types tend to have higher X-ray 
luminosities, consistent with their hosting cool cores, confirming the
trend for cool core clusters to have an early BM type discovered by
\citep{edge91b}.

\subsection*{}
While the merger hypothesis has received support from observational studies
\citep[e.g.][]{allen01b} as well as earlier simulations 
\citep[e.g.][]{ricker01,ritchie02}, the viability of cluster merging 
destroying cool cores has recently been called into question
\citep{gomez02,motl04,poole06}. The detailed investigation of \citet{poole06}
indicate that simulations of even head-on, equal-mass mergers are incapable
of disrupting an existing compact cool core, either by heating or mixing of
gas. \Citeauthor{poole06} attribute this new development in part to a
better treatment of gas cooling in their simulations, which had initially
been neglected \citep[][]{roettiger97,ricker01}. Furthermore, they
incorporate empirically motivated initial conditions for their merging
clusters \citep[outlined in][]{mccarthy04}, which result in more realistic
and compact progenitor cool cores than those in the study of
\citet{ritchie02}, for example, who nevertheless modelled the effects of
radiative cooling.

On the face of it, these new results of \citeauthor{poole06} appear to rule
out the possibility of cool cores ever being disrupted. However, this
raises the question of why cool cores are not ubiquitous, given that the
gas in all the non-CC clusters in this sample reaches cooling times as low
as a few Gyr, although here are a number of possible explanations for this
\citep[see, for example the recent review of][and references therein]{peterson06}. 
Moreover, the tendency for non-CC clusters to show signs of recent
disruption in the form of radio relics and a type II/III Bautz-Morgan
classification is not otherwise easily understood. Finally, it must be
remembered that radiative cooling and feedback are notoriously difficult
processes to model in numerical simulations, and that even the most
sophisticated schemes may fail to adequately capture subtle, small-scale
physics of the sort that could significantly affect gas dynamics in cluster
merging.

Irrespective of any causal relationship between merging activity and the
presence of a cool core, the distinctive properties of the latter reinforce
the notion of an intrinsic bi-modality within the cluster population. Such
a conclusion has important implications for the use of scaling relations as
a tool for estimating cluster properties, such as mass, since any any
bi-modality would likely dominate the scatter in these relations
\citep[see][for example]{ohara06}. The extent to which this bi-modality 
extends to the fundamental property of mass will be addressed in a
follow-up paper.

\section{Conclusions}
We have studied a statistically-selected sample of 20 clusters and groups
of galaxies, drawn from the flux-limited catalogue of \citet{ikebe02},
using data from the \Chandra\ X-ray satellite. The data have been analysed
using a non-parametric deprojection technique, to estimate the 3
dimensional temperature and density structure of the intracluster medium
(ICM). We define a quantitative method to determine the extent to which
cooling influences cluster properties: a cool core (CC) cluster is defined
as one where the ratio of the mean temperature within 0.1--0.2 \rfiveh\ to
that within 0.1\rfiveh\ exceeds unity at $>3\sigma$
significance. Accordingly we find that the sample contains 9 CC and 11
non-CC clusters.

We find that there is a clear difference between CC and non-CC clusters,
with the latter exhibiting somewhat heterogeneous properties, although
tending to be roughly isothermal within their inner regions. CC clusters,
by contrast, display a remarkable uniformity in the shape of their inner
temperature profiles. The cooling time profiles display greater uniformity
across the sample, with non-CC clusters tending to possess longer cooling
times at all radii, with a slightly flatter logarithmic
slope. Nevertheless, even non-CC clusters have cooling times much lower
than a Hubble time in all cases. This fact, together with the high
incidence of merger activity found amongst the non-CC clusters, indicates
that the gas mixing and shock heating that this entails may be responsible
for erasing cool cores or inhibiting their formation. When the gas cooling
time is plotted as a function of radius in physical units, there is a
surprising decrease in the scatter between different clusters, indicative
of a universal cooling time profile for gas in collapsed haloes. This
result suggests that radiative cooling plays a significant role in
establishing cluster properties.

\section*{Acknowledgments}
AJRS thanks Joe Mohr, Wayne Barkhouse and Alessandro Gardini for useful
discussions and suggestions. We are grateful to the referee for suggestions
which have improved the paper. AJRS acknowledges partial support from NASA
awards NNG05GI62G and GO5-6129X and from PPARC, and EO acknowledges support
from NASA award AR4-5012X. This work made use of the NASA/IPAC
Extragalactic Database (NED).

\bibliography{/data/ajrs/stuff/latex/ajrs_bibtex}

\begin{thebibliography}{}

\bibitem[\protect\citeauthoryear{{Allen}, {Fabian}, {Johnstone}, {Arnaud} \&
  {Nulsen}}{{Allen} et~al.}{2001}]{allen01b}
{Allen} S.~W.,  {Fabian} A.~C.,  {Johnstone} R.~M.,  {Arnaud} K.~A.,
  {Nulsen} P.~E.~J.,  2001, MNRAS, 322, 589

\bibitem[\protect\citeauthoryear{Allen, Schmidt \& Fabian}{Allen
  et~al.}{2002}]{allen02}
Allen S.~W.,  Schmidt R.~W.,    Fabian A.~C.,  2002, MNRAS, 334, L11

\bibitem[\protect\citeauthoryear{{Ascasibar} \& {Markevitch}}{{Ascasibar} \&
  {Markevitch}}{2006}]{ascasibar06}
{Ascasibar} Y.,  {Markevitch} M.,  2006, ApJ, submitted (astro-ph/0603246)

\bibitem[\protect\citeauthoryear{{B{\^ i}rzan}, {Rafferty}, {McNamara}, {Wise}
  \& {Nulsen}}{{B{\^ i}rzan} et~al.}{2004}]{birzan04}
{B{\^ i}rzan} L.,  {Rafferty} D.~A.,  {McNamara} B.~R.,  {Wise} M.~W.,
  {Nulsen} P.~E.~J.,  2004, ApJ, 607, 800

\bibitem[\protect\citeauthoryear{Bahcall}{Bahcall}{1977}]{bah77}
Bahcall N.~A.,  1977, ARA\&A, 15, 505

\bibitem[\protect\citeauthoryear{{Bauer}, {Fabian}, {Sanders}, {Allen} \&
  {Johnstone}}{{Bauer} et~al.}{2005}]{bauer05}
{Bauer} F.~E.,  {Fabian} A.~C.,  {Sanders} J.~S.,  {Allen} S.~W.,
  {Johnstone} R.~M.,  2005, MNRAS, 359, 1481

\bibitem[\protect\citeauthoryear{Bautz \& Morgan}{Bautz \&
  Morgan}{1970}]{bau70}
Bautz L.~P.,  Morgan W.~W.,  1970, ApJ, 162, L149

\bibitem[\protect\citeauthoryear{{Blanton}, {Sarazin}, {McNamara} \&
  {Clarke}}{{Blanton} et~al.}{2004}]{blanton04}
{Blanton} E.~L.,  {Sarazin} C.~L.,  {McNamara} B.~R.,    {Clarke} T.~E.,  2004,
  ApJ, 612, 817

\bibitem[\protect\citeauthoryear{{Buote}, {Lewis}, {Brighenti} \&
  {Mathews}}{{Buote} et~al.}{2003}]{buote03}
{Buote} D.~A.,  {Lewis} A.~D.,  {Brighenti} F.,    {Mathews} W.~G.,  2003, ApJ,
  594, 741

\bibitem[\protect\citeauthoryear{{Buote} \& {Tsai}}{{Buote} \&
  {Tsai}}{1996}]{buote96}
{Buote} D.~A.,  {Tsai} J.~C.,  1996, ApJ, 458, 27

\bibitem[\protect\citeauthoryear{Burns}{Burns}{1990}]{burns90}
Burns J.~O.,  1990, AJ, 99, 14

\bibitem[\protect\citeauthoryear{{Clarke} \& {Ensslin}}{{Clarke} \&
  {Ensslin}}{2006}]{clarke06}
{Clarke} T.~E.,  {Ensslin} T.~A.,  2006, AJ, 131, 2900

\bibitem[\protect\citeauthoryear{{Cohn} \& {White}}{{Cohn} \&
  {White}}{2005}]{cohn05}
{Cohn} J.~D.,  {White} M.,  2005, Astroparticle Physics, 24, 316

\bibitem[\protect\citeauthoryear{{Cortese}, {Gavazzi}, {Boselli},
  {Iglesias-Paramo} \& {Carrasco}}{{Cortese} et~al.}{2004}]{cortese04}
{Cortese} L.,  {Gavazzi} G.,  {Boselli} A.,  {Iglesias-Paramo} J.,
  {Carrasco} L.,  2004, A\&A, 425, 429

\bibitem[\protect\citeauthoryear{{Dickey} \& {Lockman}}{{Dickey} \&
  {Lockman}}{1990}]{dic90}
{Dickey} J.~M.,  {Lockman} F.~J.,  1990, ARA\&A, 28, 215

\bibitem[\protect\citeauthoryear{{Donahue}, {Horner}, {Cavagnolo} \&
  {Voit}}{{Donahue} et~al.}{2006}]{donahue06}
{Donahue} M.,  {Horner} D.~J.,  {Cavagnolo} K.~W.,    {Voit} G.~M.,  2006, ApJ,
  643, 730

\bibitem[\protect\citeauthoryear{{Dupke} \& {White}}{{Dupke} \&
  {White}}{2003}]{dupke03}
{Dupke} R.,  {White} R.~E.,  2003, ApJ, 583, L13

\bibitem[\protect\citeauthoryear{Edge \& Stewart}{Edge \&
  Stewart}{1991a}]{edge91b}
Edge A.~C.,  Stewart G.~C.,  1991a, MNRAS, 252, 428

\bibitem[\protect\citeauthoryear{Edge \& Stewart}{Edge \&
  Stewart}{1991b}]{edge91}
Edge A.~C.,  Stewart G.~C.,  1991b, MNRAS, 252, 414

\bibitem[\protect\citeauthoryear{Edge, Stewart \& Fabian}{Edge
  et~al.}{1992}]{edge92}
Edge A.~C.,  Stewart G.~C.,    Fabian A.~C.,  1992, MNRAS, 258, 1772

\bibitem[\protect\citeauthoryear{{Ensslin}, {Biermann}, {Klein} \&
  {Kohle}}{{Ensslin} et~al.}{1998}]{ensslin98}
{Ensslin} T.~A.,  {Biermann} P.~L.,  {Klein} U.,    {Kohle} S.,  1998, A\&A,
  332, 395

\bibitem[\protect\citeauthoryear{{Fabian}, {Nulsen} \& {Canizares}}{{Fabian}
  et~al.}{1984}]{fabian84}
{Fabian} A.~C.,  {Nulsen} P.~E.~J.,    {Canizares} C.~R.,  1984, Nature, 310,
  733

\bibitem[\protect\citeauthoryear{Finoguenov, Reiprich \&
  B{\"{o}}hringer}{Finoguenov et~al.}{2001}]{finoguenov01}
Finoguenov A.,  Reiprich T.~H.,    B{\"{o}}hringer H.,  2001, A\&A, 368, 749

\bibitem[\protect\citeauthoryear{{Fukazawa}, {Makishima} \&
  {Ohashi}}{{Fukazawa} et~al.}{2004}]{fukazawa04}
{Fukazawa} Y.,  {Makishima} K.,    {Ohashi} T.,  2004, PASJ, 56, 965

\bibitem[\protect\citeauthoryear{{Gavazzi}}{{Gavazzi}}{1978}]{gavazzi78}
{Gavazzi} G.,  1978, A\&A, 69, 355

\bibitem[\protect\citeauthoryear{{Ghizzardi}, {Molendi}, {Leccardi} \&
  {Rossetti}}{{Ghizzardi} et~al.}{2005}]{ghizzardi05}
{Ghizzardi} S.,  {Molendi} S.,  {Leccardi} A.,    {Rossetti} M.,  2005,
  preprint (astro-ph/0511445)

\bibitem[\protect\citeauthoryear{{Giovannini} \& {Feretti}}{{Giovannini} \&
  {Feretti}}{2000}]{giovannini00}
{Giovannini} G.,  {Feretti} L.,  2000, New Astronomy, 5, 335

\bibitem[\protect\citeauthoryear{{Giovannini} \& {Feretti}}{{Giovannini} \&
  {Feretti}}{2004}]{giovannini04}
{Giovannini} G.,  {Feretti} L.,  2004, Journal of Korean Astronomical Society,
  37, 323

\bibitem[\protect\citeauthoryear{{Girardi}, {Escalera}, {Fadda}, {Giuricin},
  {Mardirossian} \& {Mezzetti}}{{Girardi} et~al.}{1997}]{girardi97}
{Girardi} M.,  {Escalera} E.,  {Fadda} D.,  {Giuricin} G.,  {Mardirossian} F.,
    {Mezzetti} M.,  1997, ApJ, 482, 41

\bibitem[\protect\citeauthoryear{{G{\'o}mez}, {Loken}, {Roettiger} \&
  {Burns}}{{G{\'o}mez} et~al.}{2002}]{gomez02}
{G{\'o}mez} P.~L.,  {Loken} C.,  {Roettiger} K.,    {Burns} J.~O.,  2002, ApJ,
  569, 122

\bibitem[\protect\citeauthoryear{{Grevesse} \& {Sauval}}{{Grevesse} \&
  {Sauval}}{1998}]{grevesse98}
{Grevesse} N.,  {Sauval} A.~J.,  1998, Space Science Reviews, 85, 161

\bibitem[\protect\citeauthoryear{Helsdon \& Ponman}{Helsdon \&
  Ponman}{2000}]{hel00}
Helsdon S.~F.,  Ponman T.~J.,  2000, MNRAS, 315, 356

\bibitem[\protect\citeauthoryear{{Henriksen} \& {Tittley}}{{Henriksen} \&
  {Tittley}}{2002}]{henriksen02}
{Henriksen} M.~J.,  {Tittley} E.~R.,  2002, ApJ, 577, 701

\bibitem[\protect\citeauthoryear{{Ikebe}, {Reiprich}, {B{\" o}hringer},
  {Tanaka} \& {Kitayama}}{{Ikebe} et~al.}{2002}]{ikebe02}
{Ikebe} Y.,  {Reiprich} T.~H.,  {B{\" o}hringer} H.,  {Tanaka} Y.,
  {Kitayama} T.,  2002, A\&A, 383, 773

\bibitem[\protect\citeauthoryear{Johnstone, Allen, Fabian \& Sanders}{Johnstone
  et~al.}{2002}]{johnstone02}
Johnstone R.~M.,  Allen S.~W.,  Fabian A.~C.,    Sanders J.~S.,  2002, MNRAS,
  336, 299

\bibitem[\protect\citeauthoryear{{Kempner}, {Sarazin} \& {Ricker}}{{Kempner}
  et~al.}{2002}]{kempner02}
{Kempner} J.~C.,  {Sarazin} C.~L.,    {Ricker} P.~M.,  2002, ApJ, 579, 236

\bibitem[\protect\citeauthoryear{{Ledlow}, {Voges}, {Owen} \& {Burns}}{{Ledlow}
  et~al.}{2003}]{ledlow03}
{Ledlow} M.~J.,  {Voges} W.,  {Owen} F.~N.,    {Burns} J.~O.,  2003, AJ, 126,
  2740

\bibitem[\protect\citeauthoryear{{Leir} \& {van den Bergh}}{{Leir} \& {van den
  Bergh}}{1977}]{leir77}
{Leir} A.~A.,  {van den Bergh} S.,  1977, ApJS, 34, 381

\bibitem[\protect\citeauthoryear{{Lugger}}{{Lugger}}{1989}]{lugger89}
{Lugger} P.~M.,  1989, ApJ, 343, 572

\bibitem[\protect\citeauthoryear{Markevitch}{Markevitch}{1998}]{mar98}
Markevitch M.,  1998, ApJ, 504, 27

\bibitem[\protect\citeauthoryear{Markevitch, Forman, Sarazin \&
  Vikhlinin}{Markevitch et~al.}{1998}]{mar98b}
Markevitch M.,  Forman W.~R.,  Sarazin C.~L.,    Vikhlinin A.,  1998, ApJ, 503,
  77

\bibitem[\protect\citeauthoryear{{Markevitch}, {Vikhlinin} \&
  {Forman}}{{Markevitch} et~al.}{2003}]{markevitch03}
{Markevitch} M.,  {Vikhlinin} A.,    {Forman} W.~R.,  2003, in {Bowyer} S.,
  {Hwang} C.-Y.,  eds, Astronomical Society of the Pacific Conference Series {A
  High Resolution Picture of the Intracluster Gas}.
pp 37--+

\bibitem[\protect\citeauthoryear{{Markevitch}, {Vikhlinin} \&
  {Mazzotta}}{{Markevitch} et~al.}{2001}]{markevitch01}
{Markevitch} M.,  {Vikhlinin} A.,    {Mazzotta} P.,  2001, ApJ, 562, L153

\bibitem[\protect\citeauthoryear{{Mazzotta}, {Edge} \& {Markevitch}}{{Mazzotta}
  et~al.}{2003}]{mazzotta03}
{Mazzotta} P.,  {Edge} A.~C.,    {Markevitch} M.,  2003, ApJ, 596, 190

\bibitem[\protect\citeauthoryear{{McCarthy}, {Balogh}, {Babul}, {Poole} \&
  {Horner}}{{McCarthy} et~al.}{2004}]{mccarthy04}
{McCarthy} I.~G.,  {Balogh} M.~L.,  {Babul} A.,  {Poole} G.~B.,    {Horner}
  D.~J.,  2004, ApJ, 613, 811

\bibitem[\protect\citeauthoryear{{McLaughlin}}{{McLaughlin}}{1999}]{mcl99}
{McLaughlin} D.~E.,  1999, AJ, 117, 2398

\bibitem[\protect\citeauthoryear{{McNamara}, {Nulsen}, {Wise}, {Rafferty},
  {Carilli}, {Sarazin} \& {Blanton}}{{McNamara} et~al.}{2005}]{mcnamara05}
{McNamara} B.~R.,  {Nulsen} P.~E.~J.,  {Wise} M.~W.,  {Rafferty} D.~A.,
  {Carilli} C.,  {Sarazin} C.~L.,    {Blanton} E.~L.,  2005, Nature, 433, 45

\bibitem[\protect\citeauthoryear{{Motl}, {Burns}, {Loken}, {Norman} \&
  {Bryan}}{{Motl} et~al.}{2004}]{motl04}
{Motl} P.~M.,  {Burns} J.~O.,  {Loken} C.,  {Norman} M.~L.,    {Bryan} G.,
  2004, ApJ, 606, 635

\bibitem[\protect\citeauthoryear{{O'Hara}, {Mohr}, {Bialek} \&
  {Evrard}}{{O'Hara} et~al.}{2006}]{ohara06}
{O'Hara} T.~B.,  {Mohr} J.~J.,  {Bialek} J.~J.,    {Evrard} A.~E.,  2006, ApJ,
  639, 64

\bibitem[\protect\citeauthoryear{{O'Sullivan}, {Vrtilek} \&
  {Kempner}}{{O'Sullivan} et~al.}{2005}]{osu05b}
{O'Sullivan} E.,  {Vrtilek} J.~M.,    {Kempner} J.~C.,  2005, ApJ, 624, L77

\bibitem[\protect\citeauthoryear{{O'Sullivan}, {Vrtilek}, {Kempner}, {David} \&
  {Houck}}{{O'Sullivan} et~al.}{2005}]{osu05a}
{O'Sullivan} E.,  {Vrtilek} J.~M.,  {Kempner} J.~C.,  {David} L.~P.,    {Houck}
  J.~C.,  2005, MNRAS, 357, 1134

\bibitem[\protect\citeauthoryear{Peres, Fabian, Edge, Allen, Johnstone \&
  White}{Peres et~al.}{1998}]{peres98}
Peres C.~B.,  Fabian A.~C.,  Edge A.~C.,  Allen S.~W.,  Johnstone R.~M.,
  White D.~A.,  1998, MNRAS, 298, 416

\bibitem[\protect\citeauthoryear{{Peterson} \& {Fabian}}{{Peterson} \&
  {Fabian}}{2006}]{peterson06}
{Peterson} J.~R.,  {Fabian} A.~C.,  2006, Phys. Rep., 427, 1

\bibitem[\protect\citeauthoryear{{Piffaretti}, {Jetzer}, {Kaastra} \&
  {Tamura}}{{Piffaretti} et~al.}{2005}]{piffaretti05}
{Piffaretti} R.,  {Jetzer} P.,  {Kaastra} J.~S.,    {Tamura} T.,  2005, A\&A,
  433, 101

\bibitem[\protect\citeauthoryear{{Ponman}, {Sanderson} \&
  {Finoguenov}}{{Ponman} et~al.}{2003}]{pon03}
{Ponman} T.~J.,  {Sanderson} A.~J.~R.,    {Finoguenov} A.,  2003, MNRAS, 343,
  331

\bibitem[\protect\citeauthoryear{Poole, Fardal, Babul, McCarthy, Quinn \&
  Wadsley}{Poole et~al.}{2006}]{poole06}
Poole G.~B.,  Fardal M.~A.,  Babul A.,  McCarthy I.~G.,  Quinn T.,    Wadsley
  J.,  2006, MNRAS, submitted

\bibitem[\protect\citeauthoryear{{R Development Core Team}}{{R Development Core
  Team}}{2006}]{Rcite}
{R Development Core Team} 2006, R: A Language and Environment for Statistical
  Computing.
R Foundation for Statistical Computing, Vienna, Austria

\bibitem[\protect\citeauthoryear{{Reiprich} \& {B{\"o}hringer}}{{Reiprich} \&
  {B{\"o}hringer}}{2002}]{reiprich02}
{Reiprich} T.~H.,  {B{\"o}hringer} H.,  2002, ApJ, 567, 716

\bibitem[\protect\citeauthoryear{{Ricker} \& {Sarazin}}{{Ricker} \&
  {Sarazin}}{2001}]{ricker01}
{Ricker} P.~M.,  {Sarazin} C.~L.,  2001, ApJ, 561, 621

\bibitem[\protect\citeauthoryear{Ritchie \& Thomas}{Ritchie \&
  Thomas}{2002}]{ritchie02}
Ritchie B.~W.,  Thomas P.~A.,  2002, MNRAS, 329, 675

\bibitem[\protect\citeauthoryear{{Roettiger}, {Loken} \& {Burns}}{{Roettiger}
  et~al.}{1997}]{roettiger97}
{Roettiger} K.,  {Loken} C.,    {Burns} J.~O.,  1997, ApJS, 109, 307

\bibitem[\protect\citeauthoryear{{Rottgering}, {Snellen}, {Miley}, {de Jong},
  {Hanisch} \& {Perley}}{{Rottgering} et~al.}{1994}]{rottgering94}
{Rottgering} H.,  {Snellen} I.,  {Miley} G.,  {de Jong} J.~P.,  {Hanisch}
  R.~J.,    {Perley} R.,  1994, ApJ, 436, 654

\bibitem[\protect\citeauthoryear{{Rottgering}, {Wieringa}, {Hunstead} \&
  {Ekers}}{{Rottgering} et~al.}{1997}]{rottgering97}
{Rottgering} H.~J.~A.,  {Wieringa} M.~H.,  {Hunstead} R.~W.,    {Ekers} R.~D.,
  1997, MNRAS, 290, 577

\bibitem[\protect\citeauthoryear{{Sakelliou} \& {Ponman}}{{Sakelliou} \&
  {Ponman}}{2004}]{sakelliou04}
{Sakelliou} I.,  {Ponman} T.~J.,  2004, MNRAS, 351, 1439

\bibitem[\protect\citeauthoryear{{Sanderson}, {Finoguenov} \&
  {Mohr}}{{Sanderson} et~al.}{2005}]{san05}
{Sanderson} A.~J.~R.,  {Finoguenov} A.,    {Mohr} J.~J.,  2005, ApJ, 630, 191

\bibitem[\protect\citeauthoryear{{Sanderson} \& {Ponman}}{{Sanderson} \&
  {Ponman}}{2003}]{san03b}
{Sanderson} A.~J.~R.,  {Ponman} T.~J.,  2003, MNRAS, 345, 1241

\bibitem[\protect\citeauthoryear{{Sanderson}, {Ponman}, {Finoguenov},
  {Lloyd-Davies} \& {Markevitch}}{{Sanderson} et~al.}{2003}]{san03}
{Sanderson} A.~J.~R.,  {Ponman} T.~J.,  {Finoguenov} A.,  {Lloyd-Davies} E.~J.,
     {Markevitch} M.,  2003, MNRAS, 340, 989

\bibitem[\protect\citeauthoryear{{Slee}, {Roy}, {Murgia}, {Andernach} \&
  {Ehle}}{{Slee} et~al.}{2001}]{slee01}
{Slee} O.~B.,  {Roy} A.~L.,  {Murgia} M.,  {Andernach} H.,    {Ehle} M.,  2001,
  AJ, 122, 1172

\bibitem[\protect\citeauthoryear{{Sun}, {Jones}, {Murray}, {Allen}, {Fabian} \&
  {Edge}}{{Sun} et~al.}{2003}]{sun03}
{Sun} M.,  {Jones} C.,  {Murray} S.~S.,  {Allen} S.~W.,  {Fabian} A.~C.,
  {Edge} A.~C.,  2003, ApJ, 587, 619

\bibitem[\protect\citeauthoryear{{Sun} \& {Murray}}{{Sun} \&
  {Murray}}{2002}]{sun02}
{Sun} M.,  {Murray} S.~S.,  2002, apj, 576, 708

\bibitem[\protect\citeauthoryear{{Sun}, {Murray}, {Markevitch} \&
  {Vikhlinin}}{{Sun} et~al.}{2002}]{sun02b}
{Sun} M.,  {Murray} S.~S.,  {Markevitch} M.,    {Vikhlinin} A.,  2002, ApJ,
  565, 867

\bibitem[\protect\citeauthoryear{{Venturi}, {Bardelli}, {Zagaria}, {Prandoni}
  \& {Morganti}}{{Venturi} et~al.}{2002}]{venturi02}
{Venturi} T.,  {Bardelli} S.,  {Zagaria} M.,  {Prandoni} I.,    {Morganti} R.,
  2002, A\&A, 385, 39

\bibitem[\protect\citeauthoryear{{Vikhlinin}, {Markevitch} \&
  {Murray}}{{Vikhlinin} et~al.}{2001}]{vikhlinin01}
{Vikhlinin} A.,  {Markevitch} M.,    {Murray} S.~S.,  2001, ApJ, 551, 160

\bibitem[\protect\citeauthoryear{{Vikhlinin}, {Markevitch}, {Murray}, {Jones},
  {Forman} \& {Van Speybroeck}}{{Vikhlinin} et~al.}{2005}]{vikhlinin05}
{Vikhlinin} A.,  {Markevitch} M.,  {Murray} S.~S.,  {Jones} C.,  {Forman} W.,
   {Van Speybroeck} L.,  2005, ApJ, 628, 655

\bibitem[\protect\citeauthoryear{{Voigt} \& {Fabian}}{{Voigt} \&
  {Fabian}}{2004}]{voigt04}
{Voigt} L.~M.,  {Fabian} A.~C.,  2004, MNRAS, 347, 1130

\bibitem[\protect\citeauthoryear{{Voit}, {Balogh}, {Bower}, {Lacey} \&
  {Bryan}}{{Voit} et~al.}{2003}]{voit03b}
{Voit} G.~M.,  {Balogh} M.~L.,  {Bower} R.~G.,  {Lacey} C.~G.,    {Bryan}
  G.~L.,  2003, ApJ, 593, 272

\bibitem[\protect\citeauthoryear{{Werner}, {de Plaa}, {Kaastra}, {Vink},
  {Bleeker}, {Tamura}, {Peterson} \& {Verbunt}}{{Werner}
  et~al.}{2006}]{werner06}
{Werner} N.,  {de Plaa} J.,  {Kaastra} J.~S.,  {Vink} J.,  {Bleeker} J.~A.~M.,
  {Tamura} T.,  {Peterson} J.~R.,    {Verbunt} F.,  2006, A\&A, 449, 475

\bibitem[\protect\citeauthoryear{{Willis}, {Pacaud}, {Valtchanov}, {Pierre},
  {Ponman}, {Read}, {Andreon}, {Altieri}, {Quintana}, {Dos Santos},
  {Birkinshaw}, {Bremer}, {Duc}, {Galaz}, {Gosset}, {Jones} \&
  {Surdej}}{{Willis} et~al.}{2005}]{willis05}
{Willis} J.~P.,  {Pacaud} F.,  {Valtchanov} I.,  {Pierre} M.,  {Ponman} T.,
  {Read} A.,  {Andreon} S.,  {Altieri} B.,  {Quintana} H.,  {Dos Santos} S.,
  {Birkinshaw} M.,  {Bremer} M.,  {Duc} P.-A.,  {Galaz} G.,  {Gosset} E.,
  {Jones} L.,    {Surdej} J.,  2005, MNRAS, 363, 675

\bibitem[\protect\citeauthoryear{{Yamasaki}, {Ohashi} \& {Furusho}}{{Yamasaki}
  et~al.}{2002}]{yamasaki02}
{Yamasaki} N.~Y.,  {Ohashi} T.,    {Furusho} T.,  2002, ApJ, 578, 833

\bibitem[\protect\citeauthoryear{{Zhang}, {Boehringer}, {Finoguenov}, {Ikebe},
  {Matsushita}, {Schuecker}, {Guzzo} \& {Collins}}{{Zhang}
  et~al.}{2006}]{zhang06}
{Zhang} Y.~.,  {Boehringer} H.,  {Finoguenov} A.,  {Ikebe} Y.,  {Matsushita}
  K.,  {Schuecker} P.,  {Guzzo} L.,    {Collins} C.~A.,  2006, A\&A, accepted
  (astro-ph/0603275)

\end{thebibliography}
\label{lastpage}

\end{document}